\documentclass[12pt]{article}

\usepackage{booktabs}
\usepackage{caption}
\usepackage{subcaption}

\usepackage{siunitx}

\usepackage[table]{xcolor}
\usepackage{graphicx, amsmath} 
\usepackage[margin=1.2in]{geometry}

\usepackage[small]{titlesec}
\usepackage{tikz}
\usepackage{float}
\usetikzlibrary{arrows.meta, positioning}
\usepackage[style=numeric, sorting=none]{biblatex} 
\defbibheading{bibliography}{\section*{Bibliography}}
\definecolor{BrickRed}{RGB}{196, 30, 58} 
\definecolor{RoyalBlue}{RGB}{65, 105, 225}    

\usepackage{array}
\usepackage[hidelinks]{hyperref}

\title{\textbf{Modeling and Stabilization of Transport-Dominated Flows}\\
\vspace{3mm}
\Large{2026 Graduate Student Mathematical Modeling Camp\\ Final Report}
}

\author{%
  \begin{tabular}[t]{ccc}
    Christopher Agesen         & Sam Bradford              & Abhijnan Dikshit      \\
    {\scriptsize NJIT}         & {\scriptsize Western University} & {\scriptsize Purdue University} \\[6pt]
    Anshi Gupta                & Bhavana Morankar          & Jack Murphy           \\
    {\scriptsize U. of Houston}& {\scriptsize NC State University} & {\scriptsize U. of Arizona} \\[6pt]
    Nanda N. Raghunathan       & Karnav Raval              & Lane H. Rogers        \\
    {\scriptsize U. of Pittsburgh} & {\scriptsize Western University} & {\scriptsize U. of Tennessee} \\[12pt]
    \multicolumn{3}{c}{Mentored by Dr.\ Valeria Barra} \\
    \multicolumn{3}{c}{\scriptsize San Diego State University}
  \end{tabular}%
}

\begin{document}

\maketitle

\begin{abstract}

   \noindent
   This report explores and compares numerical stabilization methods for transport-dominated flows arising in atmospheric modeling. The Streamline-Upwind (SU) and Streamline-Upwind Petrov-Galerkin (SUPG) stabilization formulations are implemented in Julia's \texttt{ClimaCore.jl} package and Python's \texttt{Firedrake} package for a test problem with slotted-cylinder initial conditions. These methods are compared for their ability to mitigate spurious oscillations while preserving sharp features against existing hyperdiffusion and quasi-monotone limiter methods, alongside various combinations. For this benchmark, quasi-monotone limiters were most effective at minimizing un-physical extrema, at the expense of diffusing the overall structure. The SUPG method was not found to improve upon the no stabilization case, for this test case in the computed error metrics. However, it performs best at preserving sharp feature and overall structure, among tested stabilized runs. Theoretical properties of SUPG are analyzed, and an asymptotics-based SUPG algorithm is proposed as future work.
\end{abstract}

\newpage
\tableofcontents

\newpage
\section{Introduction}

Accurate numerical simulations of the Earth's atmosphere are important for studying phenomena such as climate change. Among several classical modeling challenges in atmospheric flows, the transport of a passive tracer via pure advection is especially challenging due to the nature of such problems. A pure advection problem is an example of a partial differential equation (PDE) that, when discretized with a high-order Finite Element Method (FEM) or Spectral Element Method (SEM), leads to spurious numerical oscillations and instabilities. 

As such, it is necessary to implement stabilization strategies that can mitigate oscillations and instabilities. Examples of such stabilization strategies include a fourth-order Laplacian operator (referred to as \emph{hyperdiffusion}) \cite{Ullrich2018}, slope or flux-limiters \cite{guba2014optimization}, and flux-corrected transport \cite{Zalesak1979}. Such methods, however, are not specific to the Finite or Spectral Element Methods (FEM or SEM), apply isotropic diffusion, symmetric in all directions, and often tend to overly smear out desired sharp gradients. Other methods such as the Streamline-Upwind Petrov Galerkin (SUPG) method \cite{brooks1982streamline} (and its simplified version, the Streamline-Upwind (SU) method) apply diffusion only along the streamlines of the flow and thus, can better maintain sharp gradients present in the flow. 

The main aim of this project is to systematically compare different stabilization strategies using the transport of a passive tracer as a test problem. The model of passive tracer transport (via pure advection) was implemented in the \texttt{ClimaCore.jl} \cite{ClimaCore_github} and Firedrake \cite{FiredrakeUserManual} packages, which use the Julia and Python programming languages, respectively. 

\texttt{ClimaCore.jl} \cite{ClimaCore_github} is an open-source library for the dynamical core (\emph{dycore}) of a proposed advanced Earth System Model (ESM). An ESM is a massive, multiscale, multiphysics model that couples different model components and media (including the atmosphere, ocean, and land masses) to provide an accurate representation of Earth's climate systems. The proposed ESM has been developed by the Climate Modeling Alliance (CliMA), which is a coalition of scientists, engineers, and applied mathematicians from CalTech, MIT, and the NASA Jet Propulsion Laboratory. The dynamical core (also called \emph{dycore}) of this ESM is described in this paper \cite{ClimaCore_dycore_paper}. 

In the \texttt{ClimaCore.jl} library, some stabilization strategies are readily available and utilized to counteract the numerical instability of such transport-dominated flows. These include artificial diffusion, limiters, and flux-corrected transport strategies mentioned earlier. 

In this project, the subject of primary interest is the FEM/SEM-specific SUPG \cite{brooks1982streamline} (and simplified SU method) of stabilizing numerical solutions to PDEs. We hypothesize that the SUPG method will exhibit superior performance at preserving the sharp features in transport-dominated flows (which can be of interest in problems with discontinuities or variations of scales), as this method introduces stabilization along the streamlines of the flow only---that is, anisotropically. Accordingly, we study the initial condition characterized by two high-density regions of passive tracers in the shape of slotted cylinders, as described below in Section \ref{sec:problem}. We then proceed to compare the performance of the SU and SUPG methods to alternative methods of stabilizing transport dominated flows, such as a specific class of slope/flux-limiters and hyperdiffusion in Section \ref{sec:results}. In this paper, we offer an extensive numerical and theoretical analysis of the mathematical basis of SU and SUPG method for reducing spurious oscillations and preserving sharp features in transport dominated flow problems.

In Section \ref{sec5} (Theoretical Analysis), we distinguish global conservation of density and tracer mass from pointwise boundedness of the tracer mixing ratio. We show that the conservative semidiscrete SUPG formulation preserves total mass on the closed spherical domain, but that standard linear SUPG does not guarantee nonnegativity, a discrete maximum principle, or total-variation-diminishing behavior. We then discuss residual-based discontinuity capturing and conservative flux limiting as possible modifications for suppressing crosswind oscillations and enforcing physically admissible tracer bounds. Additionally, we propose an improved time-dependent SUPG algorithm based on an asymptotic expansion of the tracer density. This algorithm allows us to solve an implicit SUPG problem by solving a series of explicit SUPG problems instead; greatly reducing computational cost.

\section{Problem Statement}
\label{sec:problem}

This section describes the slotted cylinders benchmark problem statement. This benchmark problem is similar to the test case proposed by Nair and Lauritzen \cite{nair2010class}. Consider the advection problem on the domain $\Omega$
\begin{equation}
    \begin{cases}
    \label{pure_advection}
        \frac{\partial\rho}{\partial t}=-\nabla\cdot\rho\mathbf{u}, \\
        \frac{\partial Q}{\partial t}=-\nabla\cdot Q\mathbf{u}
    \end{cases}
\end{equation}
Here, $\rho(\lambda, \theta, t) \in \Re$ is the fluid density, $Q(\lambda, \theta, t)= q(\lambda, \theta, t) * \rho (\lambda, \theta, t) \in \Re$ is the tracer density, and $\mathbf{u}[u,v]$ is the two-dimensional velocity field. The symbol $\nabla\cdot$ denotes the divergence operator on the surface of a sphere. Moreover, $\lambda, \theta, \text{and}\; t$ denote latitude, longitude, and time respectively.

For a divergent flow, $\mathbf{u}[u,v]$ is defined as
\[
\mathbf{u}[u(\lambda, \theta, t),v(\lambda, \theta, t)] = \begin{bmatrix}
u_{0}\sin^{2}(\lambda)\sin(2\theta)\cos(\frac{\pi t}{T}) + (\frac{360}{T})\cos(\theta) \\
u_{0}\sin(2\lambda)\cos(\theta)\cos(\frac{\pi t}{T})
\end{bmatrix}
\]
Here, $u_0 = \frac{2 \pi R}{T}$, with the spherical radius of the Earth $R = 6.37122 \times 10^{6}$ m and $T = 12\cdot24\cdot60\cdot60$ s represents a total duration of twelve days in seconds.

Consider the problem characterized by two slotted-cylinders for the tracer concentration as initial condition. Let $r_0 = R/2 = 3.18561 \times 10^6$ be the radii of the two slotted cylinders, centered at $(-\pi /6,0)$ and $(\pi /6,0)$ respectively. The symbols $r_{1}$ and $r_{2}$ denote the radii of the two great circles centered at these respective points. 
Accordingly, we initialize the tracer density $Q(\lambda, \theta, 0)$ as follows:
\[
Q =\begin{cases}
1, & r_1 \le r_0 \ \text{ and }\ |\lambda-\lambda_1|\,R \ge D(r_0/6),\\
1, & r_2 \le r_0 \ \text{ and }\ |\lambda-\lambda_2|\,R \ge D(r_0/6),\\
1, & r_1 \le r_0 \ \text{ and }\ |\lambda-\lambda_1|\,R < D(r_0/6)
        \ \text{ and }\ (\phi-\phi_1)\,R < D(-5r_0/12),\\
1, & r_2 \le r_0 \ \text{ and }\ |\lambda-\lambda_2|\,R < D(r_0/6)
        \ \wedge\ (\phi-\phi_2)\,R > D(5r_0/12),\\
0.1, & \text{otherwise.}
\end{cases}
\]
where $D(s)=\tfrac{180}{\pi}s$.

\subsection{SUPG Stabilization method}
The weak form of the above advection problem may be derived by multiplying the strong-form equations presented in \eqref{pure_advection} by a test function $v\in H^1(\Omega)$ to obtain 
\begin{align}
        \int_\Omega\frac{\partial\rho}{\partial t}vd\Omega
        &=-\int_\Omega\left[(\nabla\cdot\rho\mathbf{u})v\right]d\Omega, \\
        \int_\Omega\frac{\partial Q}{\partial t} \ v \ d\Omega
        &=-\int_\Omega\left[(\nabla\cdot Q \mathbf{u})v\right]d\Omega \label{eq:weak-form-for-Q}
\end{align}
In order to apply integration by parts to the above form it will be necessary to let $v$ to be in $H^1_0(\Omega)$. Since the standard Galerkin bilinear form is not coercive in the streamline direction, 
spurious oscillations tend to emerge near sharp gradients in convection-dominated problems. But an anisotropic stabilization may be achieved through the Streamline-Upwinding Petrov Galerkin (SUPG) method \cite{brooks1982streamline}. The core idea of the Petrov-Galerkin stabilization strategy is to is to replace the test function $v$ by an alternate function, $\tilde{v} = v + \tau\,\mathbf{u}\cdot\nabla v$, 
which clearly indicates the flow direction. As a part of this approach the test function resides in a different function space than the trial function, unlike the classical Galerkin formulation. Thus, the SUPG weak form of \eqref{eq:weak-form-for-Q} is given as
\[
    \int_\Omega \tilde{v}\left(\frac{\partial Q}{\partial t} + \nabla\cdot(Q\mathbf{u})\right)\mathrm{d}\Omega = 0.
\]
Upon expanding $\tilde{v}$ we get, 
\begin{align}
    \underbrace{\int_\Omega v\left(\frac{\partial Q}{\partial t} 
+ \nabla\cdot(Q\,\mathbf{u})\right)\mathrm{d}\Omega}_{\text{standard Galerkin}} 
+ \underbrace{\int_\Omega \tau\,(\mathbf{u}\cdot\nabla v)
\left(\frac{\partial Q}{\partial t} 
+ \nabla\cdot(Q\,\mathbf{u})\right)\mathrm{d}\Omega}_{\text{SUPG correction}} = 0
\end{align}

\section{Numerical Methods} \label{sec3}
In this section, we present the discrete weak formulation underlying the ClimaCore.jl and Firedrake implementations.
\subsection{SU and SUPG weak formulation: discrete implementation}
The standard Galerkin formulation for the density equation is given as follows. Find $\rho_h \in V_h \subset H^1_0(\Omega)$ such that for all
$v_h \in V_h$:
\begin{align}
    \left(\frac{\partial\rho_h}{\partial t},\, v_h\right)
  - \left(\rho_h\,\mathbf{u},\, \nabla v_h\right) = 0,
\end{align}
where $(\cdot,\cdot)$ denotes the $L^2(\Omega)$ inner product and the
divergence term has been integrated by parts. Note that SUPG correction is not applied to the density. A formulation of conservative SUPG for the tracer equation is formulated in the following lines.

\begin{table}[!b]
\centering
\caption{Description of corrections based on the residual equation used.} \label{table1}
\renewcommand{\arraystretch}{1.4}
    \begin{tabular}{lll}
      \toprule
      Mode & Residual $\mathcal{R}_{\mathrm{adv}}(q_h)$ &  \\
      \midrule
      \texttt{:SU} (Streamline Upwind only)
        & $\mathbf{u}\cdot\nabla q_h$
        & non-consistent \\
      \texttt{:consistent} (full Brooks--Hughes SUPG \cite{brooks1982streamline})
        & $\partial_t q_h + \mathbf{u}\cdot\nabla q_h$
        & consistent \\
      \bottomrule
    \end{tabular}
\end{table}

\noindent
Find $(\rho q)_h \in V_h$ such that for all $v_h \in V_h$:
\begin{equation}
\begin{aligned}
    0 &= \underbrace{
    \left(\frac{\partial Q_h}{\partial t},\, v_h\right)
    - \left(Q_h\,\mathbf{u},\, \nabla v_h\right)
    + \int_{\partial\Omega} v_h\,Q_h\,(\mathbf{u}\cdot\hat{n})\,\mathrm{d}\sigma
  }_{\text{standard Galerkin}}\\
  &\phantom{\qquad\qquad} +
  \underbrace{
    \sum_{K} \int_{\Omega_K}
    \tau\,(\mathbf{u}\cdot\nabla v_h)\,\rho_h\,\mathcal{R}_{\mathrm{adv}}(q_h)\,\mathrm{d}\Omega 
  }_{\text{SUPG correction}}.
  \label{eq:supg_weak}
\end{aligned}
\end{equation}
\noindent
Depending on the precise choice of the term $\mathcal{R}_{\mathrm{adv}}(q_h)$, the correction may be the consistent SUPG correction or the non-consistent SU correction. The various choices of $\mathcal{R}_{\mathrm{adv}}(q_h)$ and the corresponding corrections that they describe are charted in Table \ref{table1}. 
\\\\
\textbf{Petrov-Galerkin interpretation:}
To reiterate, the SU and SUPG terms may be interpreted as making use of the test function:
\begin{align}
\label{eq:supg_test_fn}
  \tilde{v}_h = v_h + \tau\,\mathbf{u}\cdot\nabla v_h,
\end{align}
so that the trial and test function spaces differ. Consequently, we perturb the test functions so that the compact support of those functions travels along the direction of the streamlines. \\

\subsection{The Stabilization Parameter $\tau$}

The code uses the Tezduyar--Osawa \cite{tezduyar2003stabilization} formula for pure
advection ($\kappa = 0$):
\begin{equation}
  \tau^{\epsilon} = \left[
    \left(\frac{2}{\Delta t}\right)^2
    + \left(\frac{2\|\mathbf{u}\|}{h}\right)^2
  \right]^{-1/2},
  \label{eq:tau_code}
\end{equation}
where $h$ is the average \emph{nodal} length scale
$h \sim h_e/p$ (element width divided by polynomial degree).\\
\textbf{Reasoning behind the stabilization parameter:}
Recall the optimal stabilisation parameter for the 1D steady
advection-diffusion problem:
\begin{align*}
    \tau^* &= \frac{h}{2\|\mathbf{u}\|}\,\xi(Pe),
  \\
  \text{where} \
  \xi(Pe) & = \coth(Pe) - \frac{1}{Pe}, \ \text{and} 
  \  Pe = \frac{\|\mathbf{u}\|\,h}{2\varepsilon}.
\end{align*}
But for an advection dominated problem($Pe \gg 1$, $\varepsilon \to 0$), 
\[
  \coth(Pe) \to 1, \qquad \frac{1}{Pe} \to 0
  \qquad\Rightarrow\qquad \xi(Pe) \to 1,
\]
and hence,
\[
  \tau^* \;\longrightarrow\; \frac{h}{2\|\mathbf{u}\|}.
\]
\textbf{The UGN element length scale:}
{UGN} stands for {U}pwind-direction, {G}lobal,
{N}odal, describing how the element length scale
$h_{\mathrm{UGN}}$ is constructed \cite{tezduyar2003stabilization}:
\begin{itemize}
  \item \textbf{U}: the length is measured in the \emph{upwind
    (streamline) direction} $\hat{\mathbf{u}} = \mathbf{u}/\|\mathbf{u}\|$,
    not isotropically.
  \item \textbf{G}: it is a \emph{global} element length, using
    \emph{all} nodes of the element by summing over all basis functions
    $N_a$.
  \item \textbf{N}: it is \emph{nodal}, computed directly from the
    nodal basis functions $N_a$ and their gradients $\nabla N_a$.
\end{itemize}
The explicit formula from \cite{tezduyar2003stabilization} is:
\[
  h_{\mathrm{UGN}}
  = 2\|\mathbf{u}\|
    \left(\sum_{a=1}^{n_{\mathrm{en}}}
    \bigl|\mathbf{u}\cdot\nabla N_a\bigr|\right)^{-1},
\]
where $n_{\mathrm{en}}$ is the number of nodes per element. Geometrically, $h_{\mathrm{UGN}}$ is the distance spanned by $\mathbf{u}$ across the element in the streamline direction.
{SUGN} stands for {S}tabilisation parameter,
{U}pwind, {G}lobal, {N}odal --- it is simply the
prefix {S} (stabilisation parameter) appended to UGN. The three components are \cite{tezduyar2003stabilization}:

\begin{align*}
  \tau_{\mathrm{SUGN1}} &= \frac{h_{\mathrm{UGN}}}{2\|\mathbf{u}\|}
  && \text{(advective timescale)},\\[4pt]
  \tau_{\mathrm{SUGN2}} &= \frac{\Delta t}{2}
  && \text{(temporal timescale)},\\[4pt]
  \tau_{\mathrm{SUGN3}} &= \frac{h_{\mathrm{UGN}}^2}{4\nu}
  && \text{(diffusive timescale)}.
\end{align*}

\noindent
They are combined as an $\ell^2$ inverse norm:
\[
  (\tau_{\mathrm{SUPG}})_{\mathrm{UGN}}
  = \left[
      \frac{1}{\tau_{\mathrm{SUGN1}}^2}
    + \frac{1}{\tau_{\mathrm{SUGN2}}^2}
    + \frac{1}{\tau_{\mathrm{SUGN3}}^2}
    \right]^{-1/2},
\]
so that $\tau$ is always bounded by the smallest of the three
timescales.
By removing the diffusing term, thus we get the formula defined in \eqref{eq:tau_code}.

\subsection{Implementation details for the ClimaCore.jl code}
The ClimaCore.jl code used in testing was created by augmenting the existing sphere example \texttt{limiters\_advection.jl} in the public repository \cite{ClimaCore_github} (located in \\ \texttt{ClimaCore.jl/examples/sphere/limiters\_advection.jl}). The quasi-monotone limiters and hyperdiffusion stabilization techniques were already implemented in the ClimaCore.jl API and exercised in the \texttt{limiters\_advection.jl} example file. SU and SUPG were novel implementations, developed in this project. After computing the weak form of the solution, the resulting differential equation is solved using the tri-stage, third-order Strong Stability Preserving Runge-Kutta method (SSPRK33) from the \texttt{OrdinaryDiffEq.jl} package. At each time step, a weighted direct stiffness summation is applied for continuity enforcement of the continuous-Galerkin (CG) spectral elements. Table \ref{table2} below highlights choices made for key solver and discretization parameters in the code when solving the test case considered in this work.

\begin{table}[!b]
\centering
\caption{Discretization and solver parameters used in the ClimaCore.jl code.} \label{table2}
\renewcommand{\arraystretch}{1.4}
    \begin{tabular}{ll}
      \toprule
      Parameter & Value  \\
      \midrule
      Number of horizontal elements per cube panel 
        &  20 \\
      Number of quadrature nodes
        &  4 \\
      Time step size, $\Delta t$
        &  345.60 s\\
      Number of time steps
        &  3000\\
      Hyperdiffusion coefficient
        &  $6.6\times 10^{14}$\\
      \bottomrule
    \end{tabular}
\end{table}

The example code enables the combination of different stabilization methods. SU and SUPG cannot be applied in the same run, but either can be paired with limiters hyperdiffusion, or both limiters and hyperdiffusion. The implementation of the continuous equations and residual calculation for SUPG is described below. 

Let $\Omega $ be the sphere. The state consists of density $\rho$ and
tracer $q$ (e.g.\ concentration). The strong form of the
system is:
\begin{align}
  \frac{\partial \rho}{\partial t} + \nabla \cdot (\rho\,\mathbf{u}) &= 0
  \label{eq:density} \\[6pt]
  \frac{\partial (\rho q)}{\partial t} + \nabla \cdot (\rho q\,\mathbf{u}) &= 0
  \label{eq:tracer}
\end{align}
where $\mathbf{u}$ is a prescribed, time-dependent velocity field. Because the continuity equation \eqref{eq:density} holds, the tracer equation \eqref{eq:tracer} can be rewritten via the product rule:
\[
  \frac{\partial(\rho q)}{\partial t} + \nabla\cdot(\rho q\,\mathbf{u})
  = \rho\underbrace{\left(\frac{\partial q}{\partial t}
    + \mathbf{u}\cdot\nabla q\right)}_{\mathcal{R}_{\mathrm{adv}}(q)}
  + q\underbrace{\left(\frac{\partial\rho}{\partial t}
    + \nabla\cdot(\rho\,\mathbf{u})\right)}_{=\,0\text{ by \eqref{eq:density}}}
  = \rho\,\mathcal{R}_{\mathrm{adv}}(q).
\]
Hence at the \emph{continuous} level the conservative residual of
\eqref{eq:tracer} factors as:
\[
  {R(\rho q) = \rho\,\mathcal{R}_{\mathrm{adv}}(q), \qquad  \text{where}
  \qquad \mathcal{R}_{\mathrm{adv}}(q)
  = \frac{\partial q}{\partial t} + \mathbf{u}\cdot\nabla q.}
\]
At the discrete level, if $\rho_h$ does not satisfy \eqref{eq:density} exactly (which it generally will not), then the conservative and advective forms decouple.


\subsection{Implementation details for the Firedrake code}

\noindent The Firedrake library \cite{FiredrakeUserManual} implementation utilized the SU stabilization method for solving the advection problem. Originally, the SUPG stabilization implemented in the Gusto library \cite{gusto_software} was utilized. However, further exploration uncovered that stabilization is only implemented in Gusto for planar meshes rather than spherical meshes, which are of interest in this work, since the pure advection problem is modeled on a cubed spherical mesh in this project. As such, the SU stabilization was directly implemented through the Firedrake library using the weak form of the governing equations. The Gusto library example for the slotted cylinders benchmark \cite{gusto_software} was adapted to test the Firedrake implementation. 

Firedrake allows the user to directly specify the test function space (whereas this is indirectly implemented in ClimaCore.jl and not part of the public-facing API). This allowed an easy specification of the SU correction test function shown in (\ref{eq:supg_test_fn}). This test function can then be used to specify the linear and bilinear forms of FEM as follows.

\noindent Casting $\partial Q/\partial t + \mathbf{u}\cdot\nabla Q = 0$ as a method-of-lines problem for the rate $k \equiv \partial Q/\partial t$, each stage solves: find $k$ with
\begin{equation}
\underbrace{\int_\Omega v_h\, k \, dx}_{a(w_h,k)} = \underbrace{-\int_\Omega \tilde{v}_h\,(\mathbf{u}\cdot\nabla Q)\, dx}_{L(w_h)}, \quad \tilde{v}_h = v_h + \tau\,\mathbf{u}\cdot\nabla v_h, \quad \forall\, v_h \in V_h.
\label{eq:weak_form}
\end{equation}
Since the SU correction is applied to the advection term only and not the temporal term, $a(\phi,k)$ is the standard Galerkin mass matrix, independent of $\mathbf{u}$ and $\tau$. It is therefore assembled and factorised once (\texttt{constant\_jacobian=True}) and reused across all stages, unlike the full SUPG, where the mass matrix must be reassembled whenever $\mathbf{u}$ changes; the trade-off is that the stabilization is only $\mathcal{O}(h)$ consistent. The stabilization parameter follows the Tezduyar formula as given in (\ref{eq:tau_code}). To calculate the stabilization parameter, $h = \sqrt{|K|}$ is computed from the local cell area, $|K|$, and stored as a $DG0$ field in Firedrake. 


Subsequently, the \texttt{LinearVariationalSolve} functionality from Firedrake can be used to solve the weak form of the equations. To advance the solution in time, the SSPRK3 scheme \cite{shu1988efficient} is implemented, which solves the weak form of the equations and updates the solution at each time step. In this way, the Firedrake implementation can apply the SU stabilization to an unsteady advection problem.

\noindent With $k(D,t)$ the rate from (\ref{eq:weak_form}) at field $D$ and velocity time $t$, the step $D^{(n)} \to D^{(n+1)}$ is
\begin{align}
D^{(1)} &= D^{(n)} + \Delta t\, k\left(D^{(n)},\, t^{(n)}\right), \nonumber\\
D^{(2)} &= \tfrac{3}{4} D^{(n)} + \tfrac{1}{4} D^{(1)} + \tfrac{1}{4}\Delta t\, k\left(D^{(1)},\, t^{(n)} + \Delta t\right), \label{eq:rk}\\
D^{(n+1)} &= \tfrac{1}{3} D^{(n)} + \tfrac{2}{3} D^{(2)} + \tfrac{2}{3}\Delta t\, k\left(D^{(2)},\, t^{(n)} + \tfrac{1}{2}\Delta t\right), \nonumber
\end{align}

\noindent with the velocity re-evaluated at the stage abscissae $c = [0,\, 1,\, \tfrac{1}{2}]$. Each stage is one LU solve of the pre-factored mass matrix; being strong-stability-preserving, SSPRK3 adds no oscillations beyond those in the spatially stabilized operator. Table \ref{table3} below details choices for important discretization and solver parameters used in the Firedrake code. The refinement level is an indication of the spatial grid size. A refinement level of $n$ indicates that there are $2^n$ segments along each edge of a panel in the discretization. 

\begin{table}[!b]
\centering
\caption{Discretization and solver parameters used in the Firedrake code.} \label{table3}
\renewcommand{\arraystretch}{1.4}
    \begin{tabular}{ll}
      \toprule
      Parameter & Value  \\
      \midrule
      Refinement level &  4\\
      Time step size, $\Delta t$ & 345.60 s \\
      Number of time steps & 3000 \\
      Polynomial degree of elements &  3\\
      Quadrature degree & 6\\
      \bottomrule
    \end{tabular}
\end{table}

\section{Results and Discussion} \label{sec:results}

We now compare numerical solutions of the passive-tracer advection system \eqref{pure_advection} under various stabilization (and combinations thereof) methods, using both the \texttt{ClimaCore.jl} \cite{ClimaCore_github} and Firedrake \cite{FiredrakeUserManual} packages. We compare the simulations both qualitatively through the tracer fields and quantitatively using particular error metrics of interest. 

\begin{figure}[htbp!]
    \centering
    \includegraphics[trim={0cm 1cm 0cm 2cm}, clip, width=0.8\linewidth]{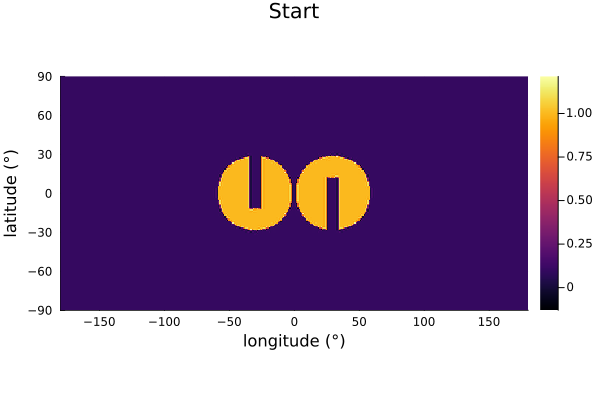}
    \caption{Initial condition for all simulations showing two slotted cylinders, plotted on a latitude-longitude grid.}
    \label{fig:initialcondition}
\end{figure}

We solve the advection system with the slotted-cylinder test case, with the initial condition as shown in Figure \ref{fig:initialcondition}. Following the examples in \texttt{ClimaCore.jl} \cite{ClimaCore_github}, this is chosen such that the tracer field returns to its initial configuration after one cycle $t \in [0, T]$. 
As such, we choose two classes of metrics to quantify the error between the initial and final field. In order to evaluate if the tracer field remains physically bounded (no spurious undershoots/overshoots), we compute
\begin{align}
    q_{\mathrm{over}}
    &=
    \frac{
        \max_{\lambda,\theta} q_T
        -
        \max_{\lambda,\theta} q_0
    }{\Delta q_0},
    &
    q_{\mathrm{under}}
    &=
    \frac{
        \min_{\lambda,\theta} q_T
        -
        \min_{\lambda,\theta} q_0
    }{\Delta q_0},
    \label{eq:extrema_errors}
\end{align}
where $q_0$ and $q_T$ denote the initial and final tracer fields (respectively) and the initial tracer range is defined as
\begin{equation}
    \Delta q_0
    =
    \max_{\lambda,\theta} q_0
    -
    \min_{\lambda,\theta} q_0.
\end{equation}
\noindent
While \eqref{eq:extrema_errors} measure the appearance of local nonphysical extrema, they do not measure the preservation of the overall shape of the tracer field. Thus, we also compute the global errors between the initial and final tracer fields
\begin{align}
    \ell_1
    =
    \frac{I\!\left[|q_T-q_0|\right]}
         {I\!\left[|q_0|\right]}, \ \ 
    \ell_2
    =
    \left(
    \frac{I\!\left[(q_T-q_0)^2\right]}
         {I\!\left[q_0^2\right]}
    \right)^{1/2}, \ \ 
    \ell_\infty
    =
    \frac{
        \max_{\lambda,\theta}|q_T-q_0|
    }{
        \max_{\lambda,\theta}|q_0|
    },
    \label{eq:global_errors}
\end{align}
where 
\begin{align}
    I[f]
    =
    \frac{1}{4\pi}
    \int_0^{2\pi}
    \int_{-\pi/2}^{\pi/2}
    f(\lambda,\theta)\cos\theta\,
    \mathrm{d}\theta\,\mathrm{d}\lambda
\end{align}
denotes the spherical area average. 

We numerically solve the passive-tracer advection system \eqref{pure_advection} for the slotted-cylinder test case, using hyperdiffusion, quasi-monotone limiting, SU, SUPG, and selected combinations of these stabilization methods. The error metrics \eqref{eq:extrema_errors} and \eqref{eq:global_errors} are computed for each case, and shown in Table \ref{tab:stabilization-comparison}. 

%

\begin{table}[htbp!]
\centering
\caption{Comparison of simulations with different stabilization choices, including combinations. All results were computed using ClimaCore.jl except for the labeled Firedrake run. Here H, L, SU, and SUPG denote hyperdiffusion, the quasi-monotone limiter, streamline upwind, and streamline-upwind Petrov--Galerkin stabilization, respectively.}

\label{tab:stabilization-comparison}
\small
\setlength{\tabcolsep}{6pt}
\renewcommand{\arraystretch}{1.2}

\begin{tabular}{
    l
    S[table-format=1.2e-2]
    S[table-format=-1.2e-2]
    S[table-format=1.2e-2]
    S[table-format=1.2e-2]
    S[table-format=1.2e-2]
}
\toprule
\textbf{Stabilization}
& {$q_{\mathrm{over}}$}
& {$q_{\mathrm{under}}$}
& {$\ell_1$}
& {$\ell_2$}
& {$\ell_\infty$} \\
\midrule
None & 9.43e-3 & -1.37e-02 & 9.60e-04 & 4.39e-03 & 1.34e-01 \\
H & 1.24e-01 & -1.42e-01 & 1.62e-01 & 5.99e-01 & 6.82e0 \\
L & 4.69e-07 & -7.32e-08 & 9.86e-02 & 4.48e-01 & 6.30e0 \\
SU & 3.99e-02 & -7.36e-02 & 9.39e-02 & 4.36e-01 & 4.50 \\
SU (Firedrake) & 6.57e-02 & -7.83e-02 & 7.92e-02 & 1.70e-01 & 6.17e-01 \\
SUPG & 5.96e-02 & -4.61e-02 & 1.43e-02 & 2.75e-02 & 6.25e-01 \\
H + L & 1.41e-07 & -7.32e-08 & 1.43e-01 & 5.98e-01 & 7.01 \\
L + SU & 4.69e-07 & -7.32e-08 & 1.21e-01 & 5.14e-01 & 5.52 \\
L + SUPG & 4.69e-07 & -7.32e-08 & 9.86e-02 & 4.48e-01 & 6.30 \\
H + L + SU & 1.38e-07 & -7.32e-08 & 1.63e-01 & 6.34e-01 & 6.58 \\
H + L + SUPG & 1.41e-07 & -7.32e-08 & 1.43e-01 & 5.98e-01 & 7.01 \\  
\bottomrule
\end{tabular}
\end{table}

\begin{figure}[htbp!]
    \centering

    \begin{subfigure}[b]{0.48\textwidth}
        \centering
        \includegraphics[trim={0 0 0 6cm}, clip, width=\linewidth]{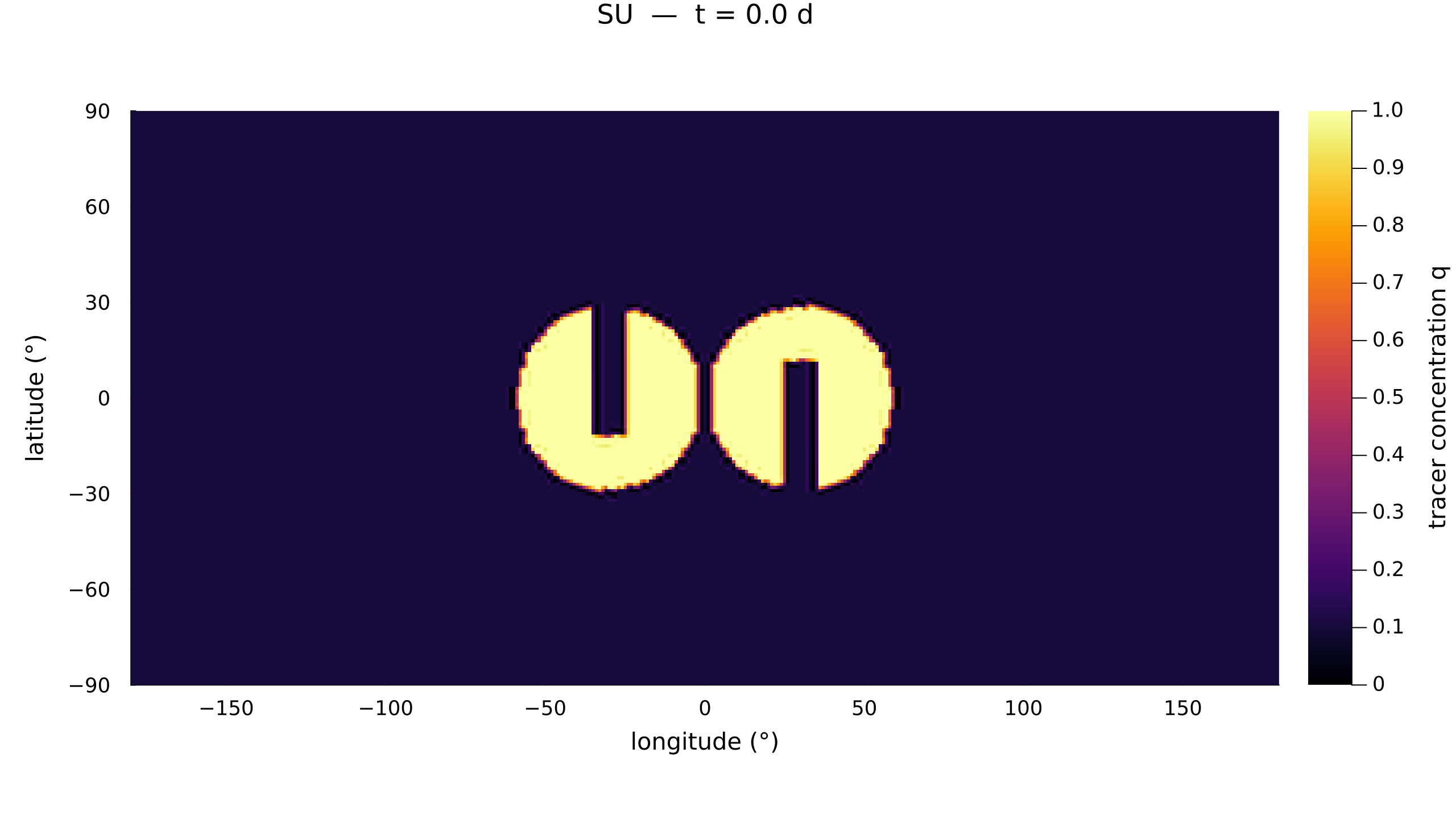}
        \caption{ClimaCore.jl, $t=0$.}
    \end{subfigure}
    \hfill
    \begin{subfigure}[b]{0.48\textwidth}
        \centering
        \includegraphics[width=\linewidth]{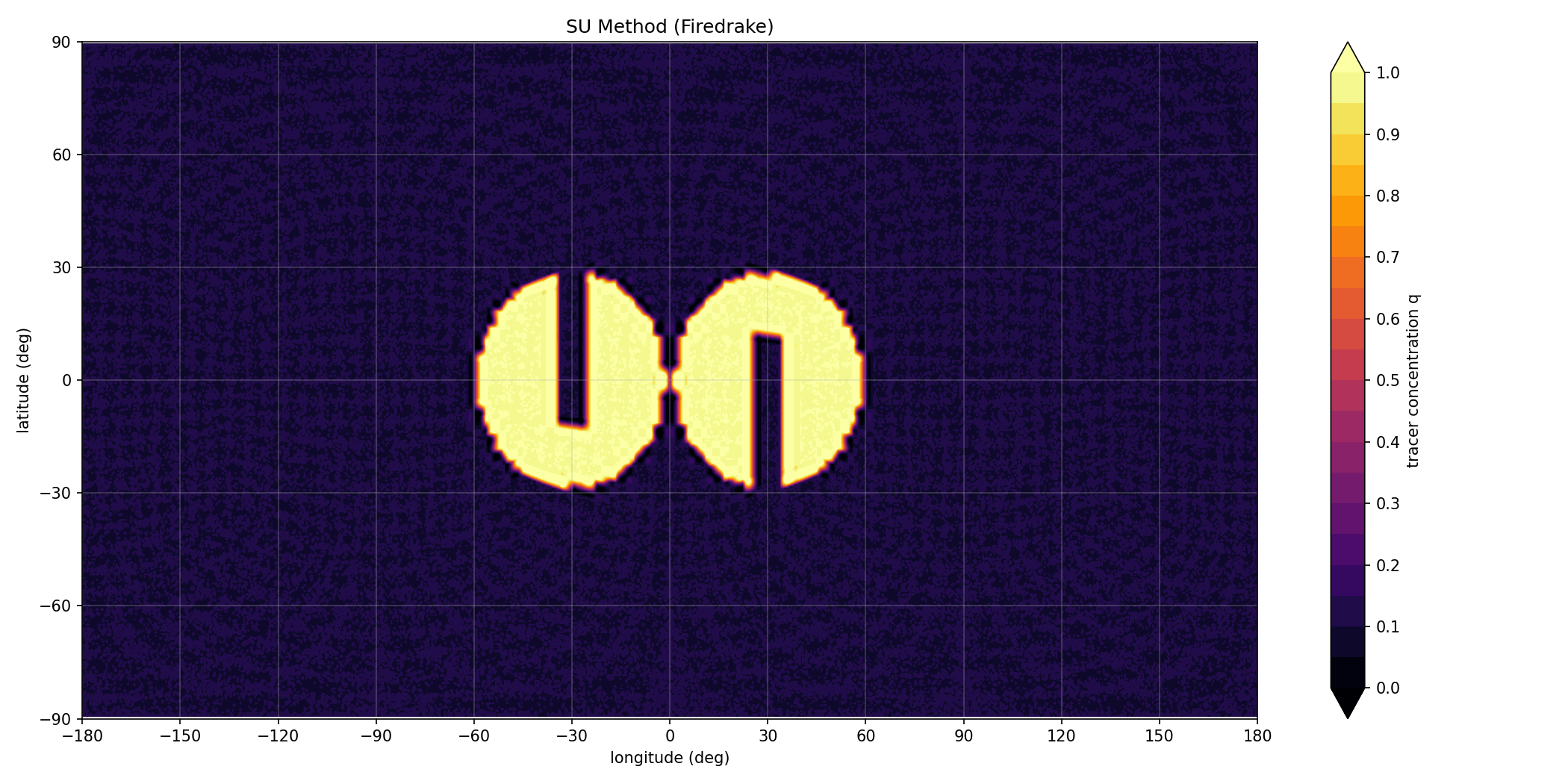}
        \caption{Firedrake, $t=0$.}
    \end{subfigure}

    \vspace{0.8em}

    \begin{subfigure}[b]{0.48\textwidth}
        \centering
        \includegraphics[trim={0 0 0 6cm}, clip, width=\linewidth]{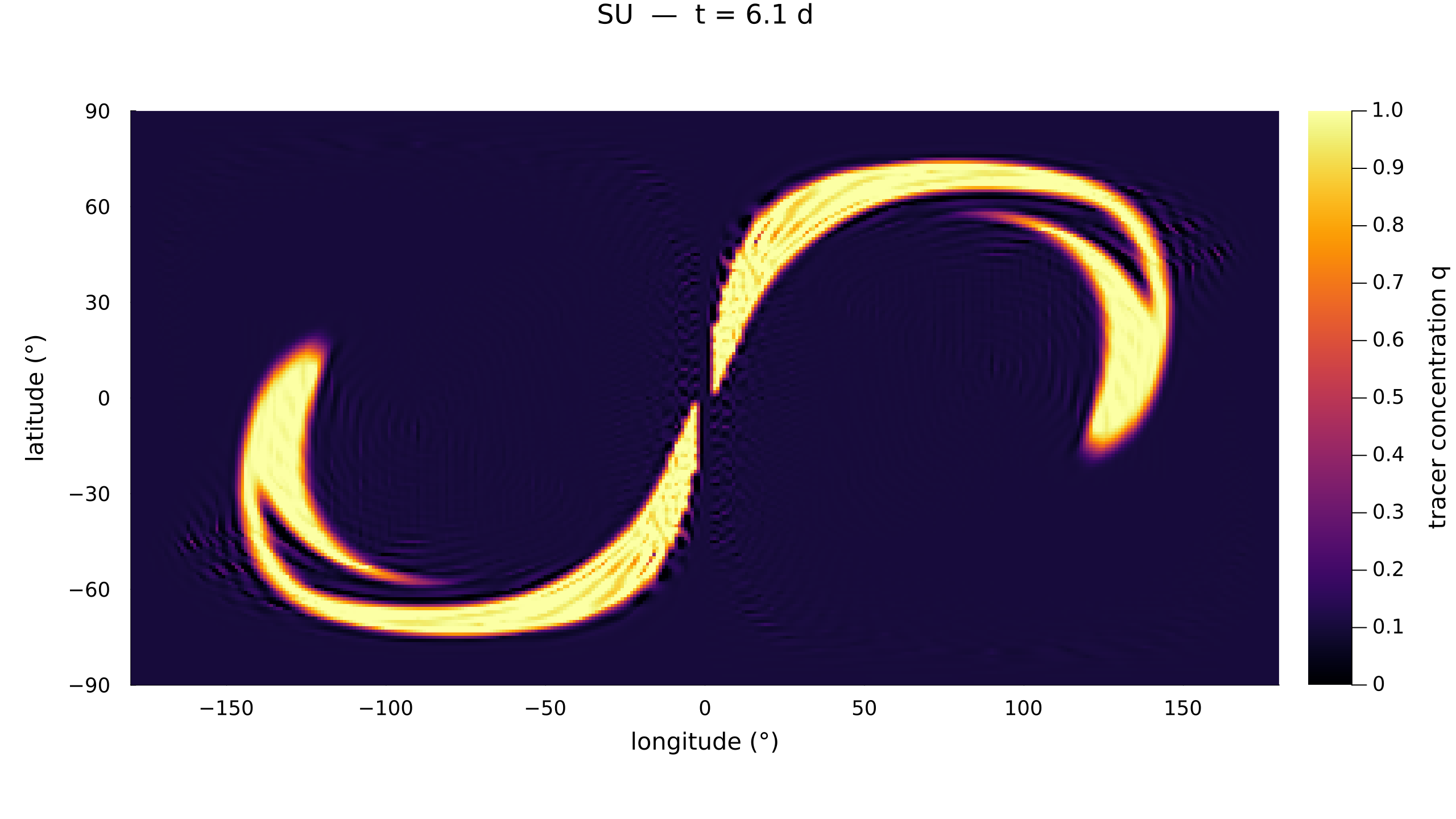}
        \caption{ClimaCore.jl, $t=T/2$.}
    \end{subfigure}
    \hfill
    \begin{subfigure}[b]{0.48\textwidth}
        \centering
        \includegraphics[width=\linewidth]{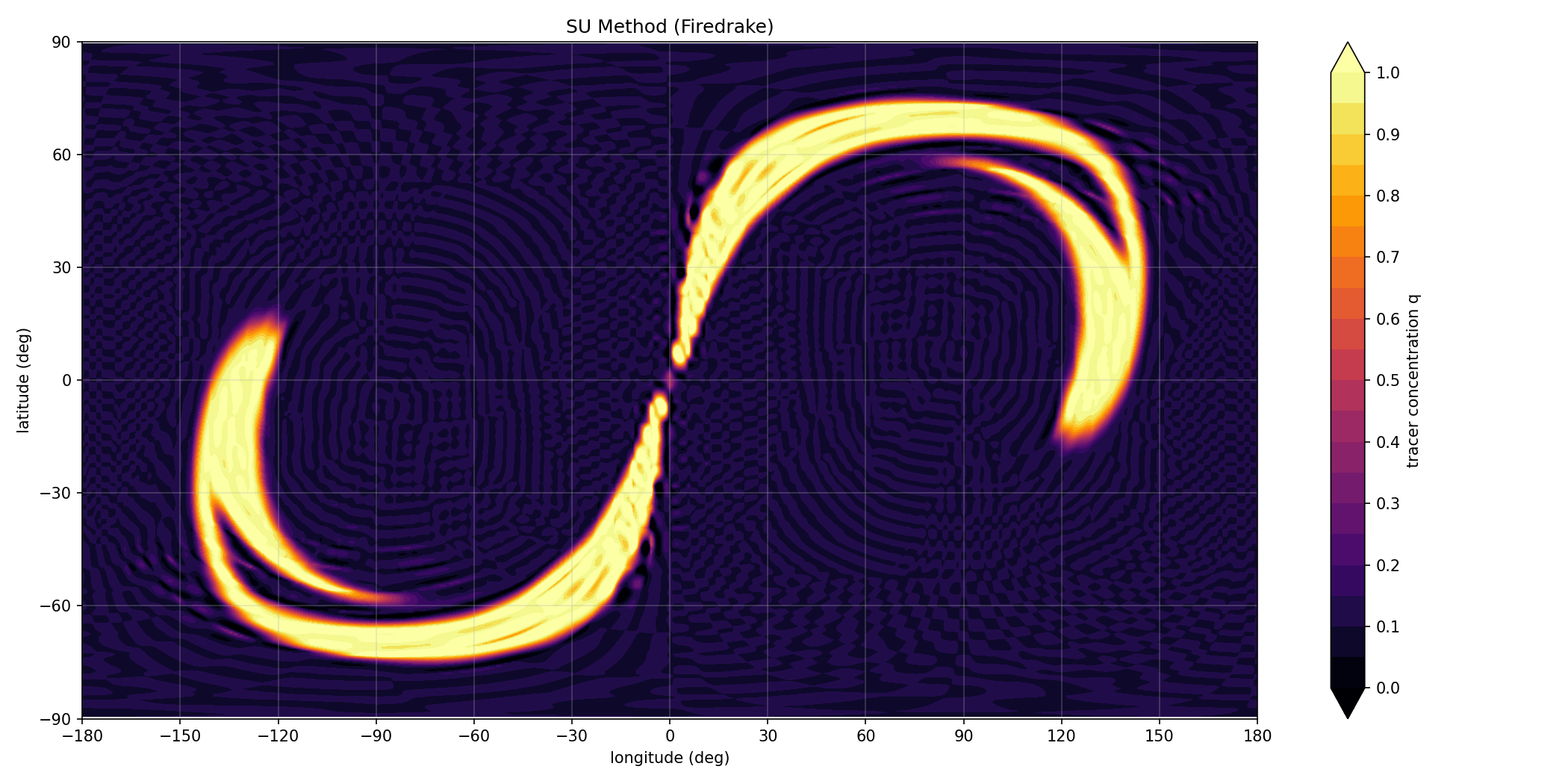}
        \caption{Firedrake, $t=T/2$.}
    \end{subfigure}

    \vspace{0.8em}

    \begin{subfigure}[b]{0.48\textwidth}
        \centering
        \includegraphics[trim={0 0 0 6cm}, clip, width=\linewidth]{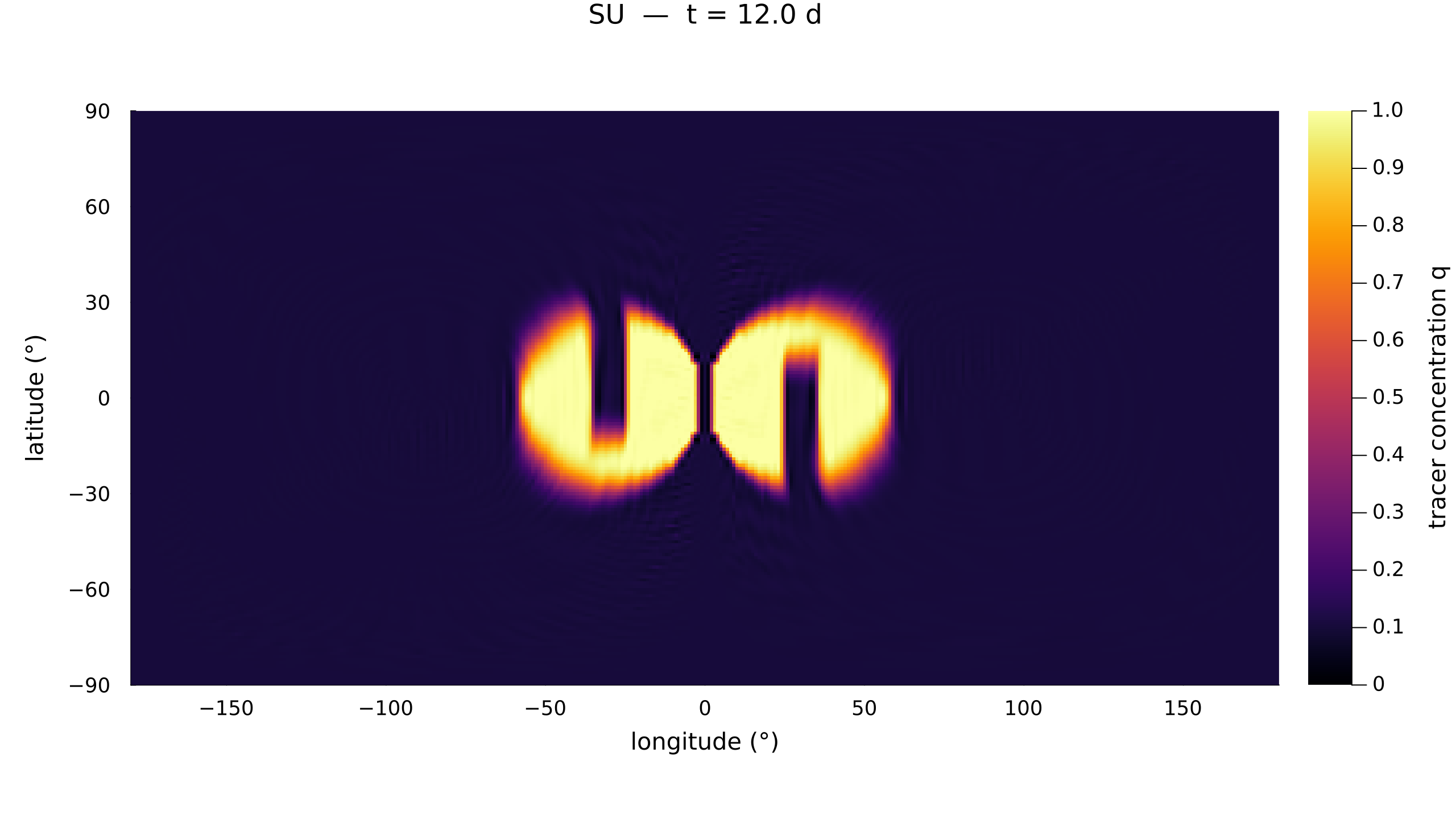}
        \caption{ClimaCore.jl, $t=T$.}
    \end{subfigure}
    \hfill
    \begin{subfigure}[b]{0.48\textwidth}
        \centering
        \includegraphics[width=\linewidth]{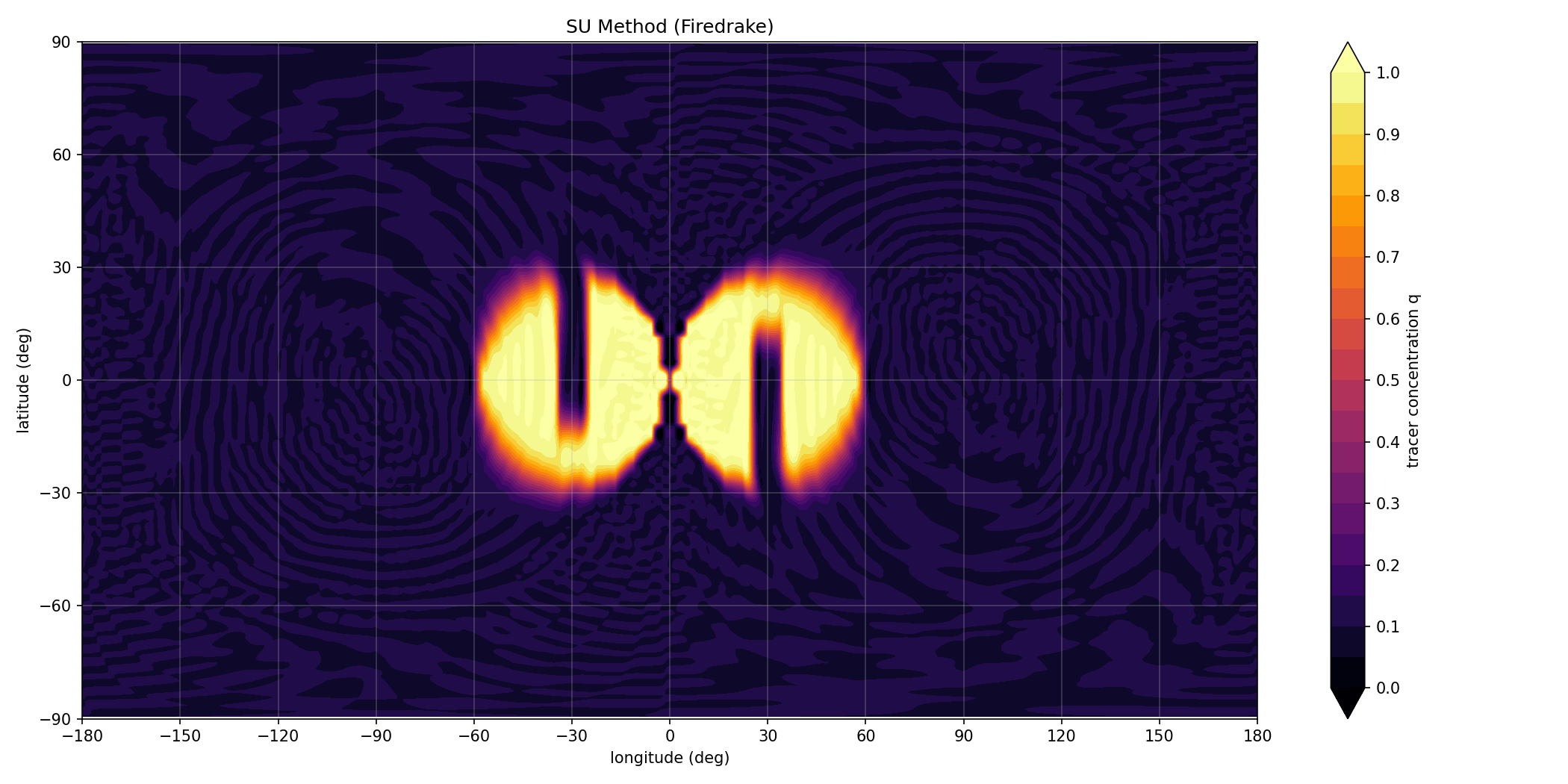}
        \caption{Firedrake, $t=T$.}
    \end{subfigure}

    \caption{Comparison of the streamline-upwind (SU) solutions computed
    with ClimaCore.jl and Firedrake. The top row confirms the common initial
    tracer configuration at $t=0$; the middle and bottom rows show the
    tracer fields at $t=T/2$ and $t=T$, respectively.}
    \label{fig:su-code-comparison}
\end{figure}

First, we turn our attention to the solutions computed using the SU stabilization method, as this was the case computed by both the ClimaCore.jl and Firedrake implementations. The solutions at the initial, intermediate ($t = T/2$) and final ($t = T$) times are shown in Figure \ref{fig:su-code-comparison}. Observe that the solutions computed using different implementations qualitatively match with one another over the time domain, with visibly comparable smearing of the sharp tracer features by the final time. This is also reflected in the error results shown in Table \ref{tab:stabilization-comparison}, where all of the metrics match up to the order of magnitude, except for $\ell_\infty$. This comparison provides a useful consistency check, as the main behavior of the SU method matches between two independent implementations. However, the discrepancy in $\ell_\infty$ indicates that they are not numerically identical, and more work is required to find the source of this difference. 


\begin{figure}[htbp!]
    \centering

    \begin{subfigure}[b]{0.48\textwidth}
        \centering
        \includegraphics[width=\linewidth]{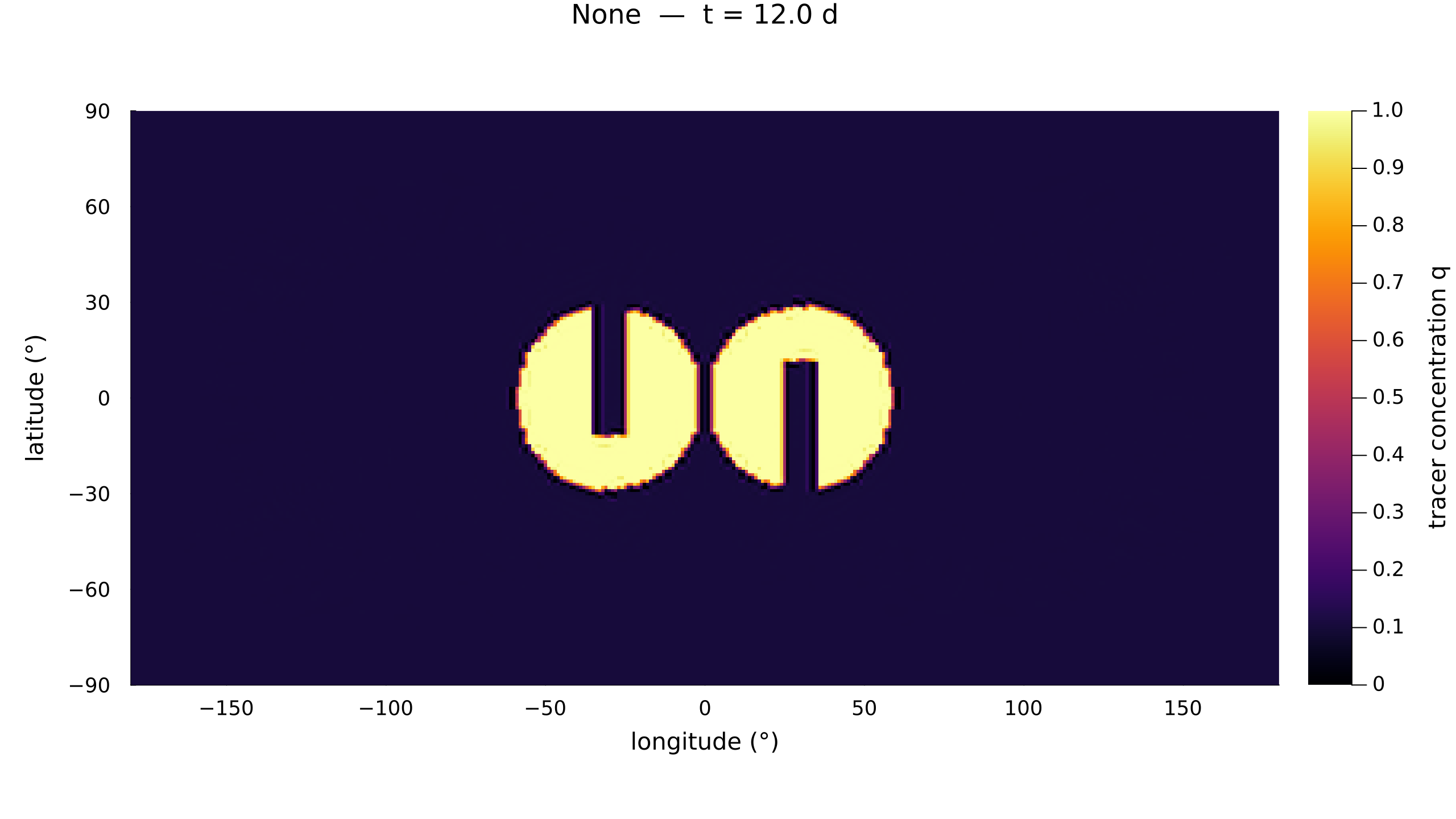}
        \caption{No stabilization.}
        \label{fig:none-final}
    \end{subfigure}
    \hfill
    \begin{subfigure}[b]{0.48\textwidth}
        \centering
        \includegraphics[width=\linewidth]{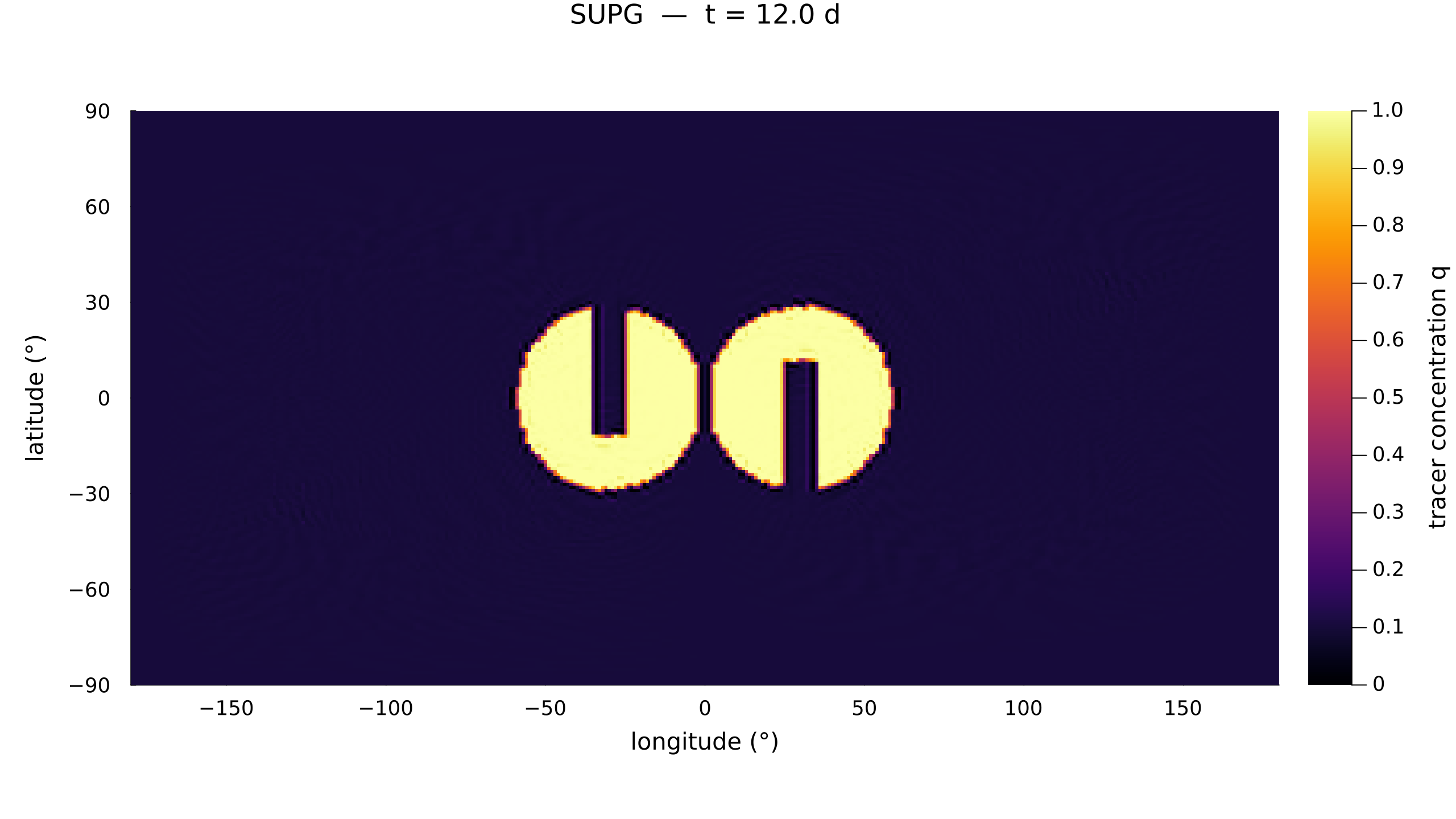}
        \caption{SUPG.}
        \label{fig:supg-final}
    \end{subfigure}

    \vspace{0.8em}

    \begin{subfigure}[b]{0.48\textwidth}
        \centering
        \includegraphics[width=\linewidth]{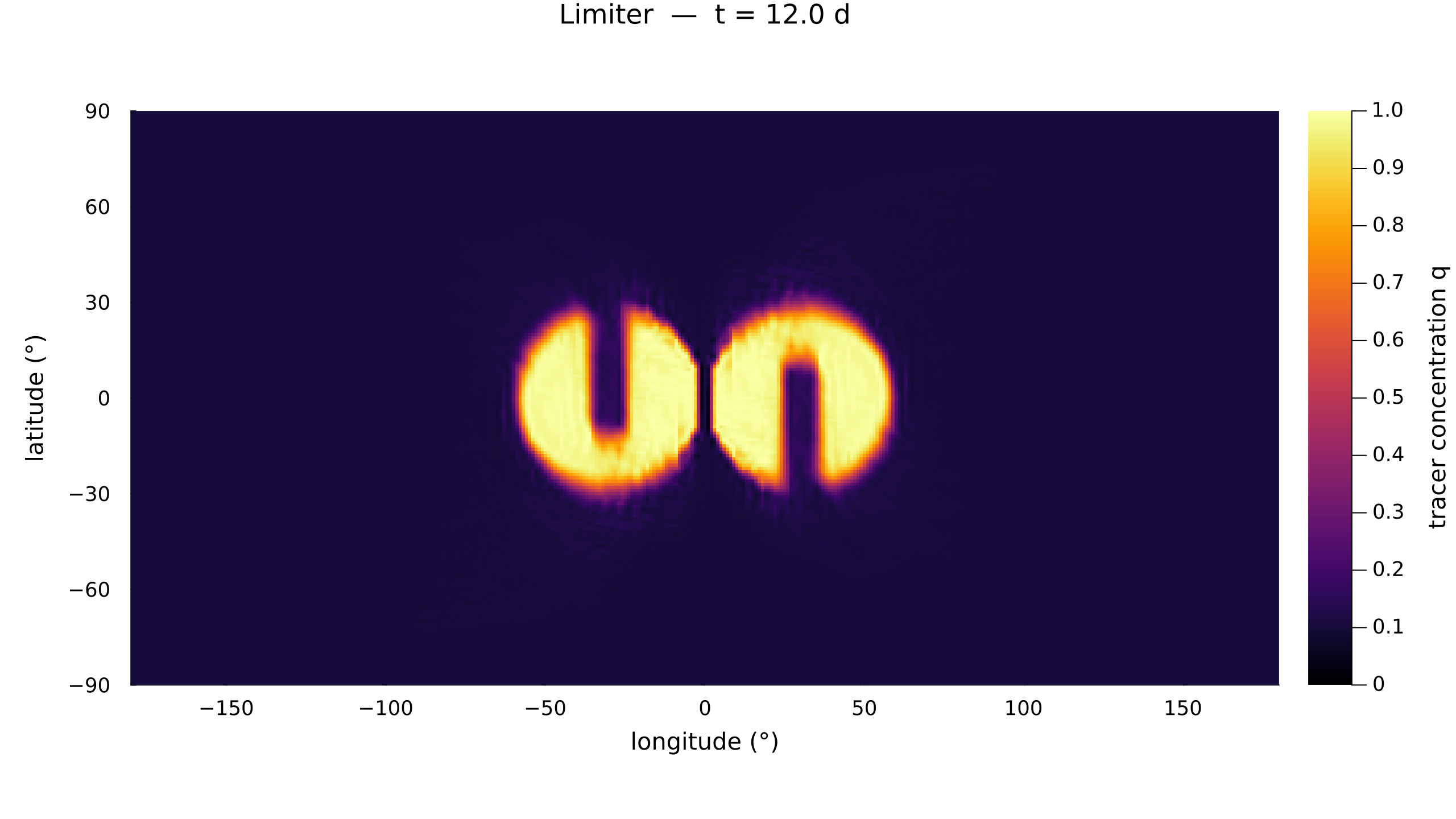}
        \caption{Quasi-monotone limiter.}
        \label{fig:limiter-final}
    \end{subfigure}
    \hfill
    \begin{subfigure}[b]{0.48\textwidth}
        \centering
        \includegraphics[width=\linewidth]{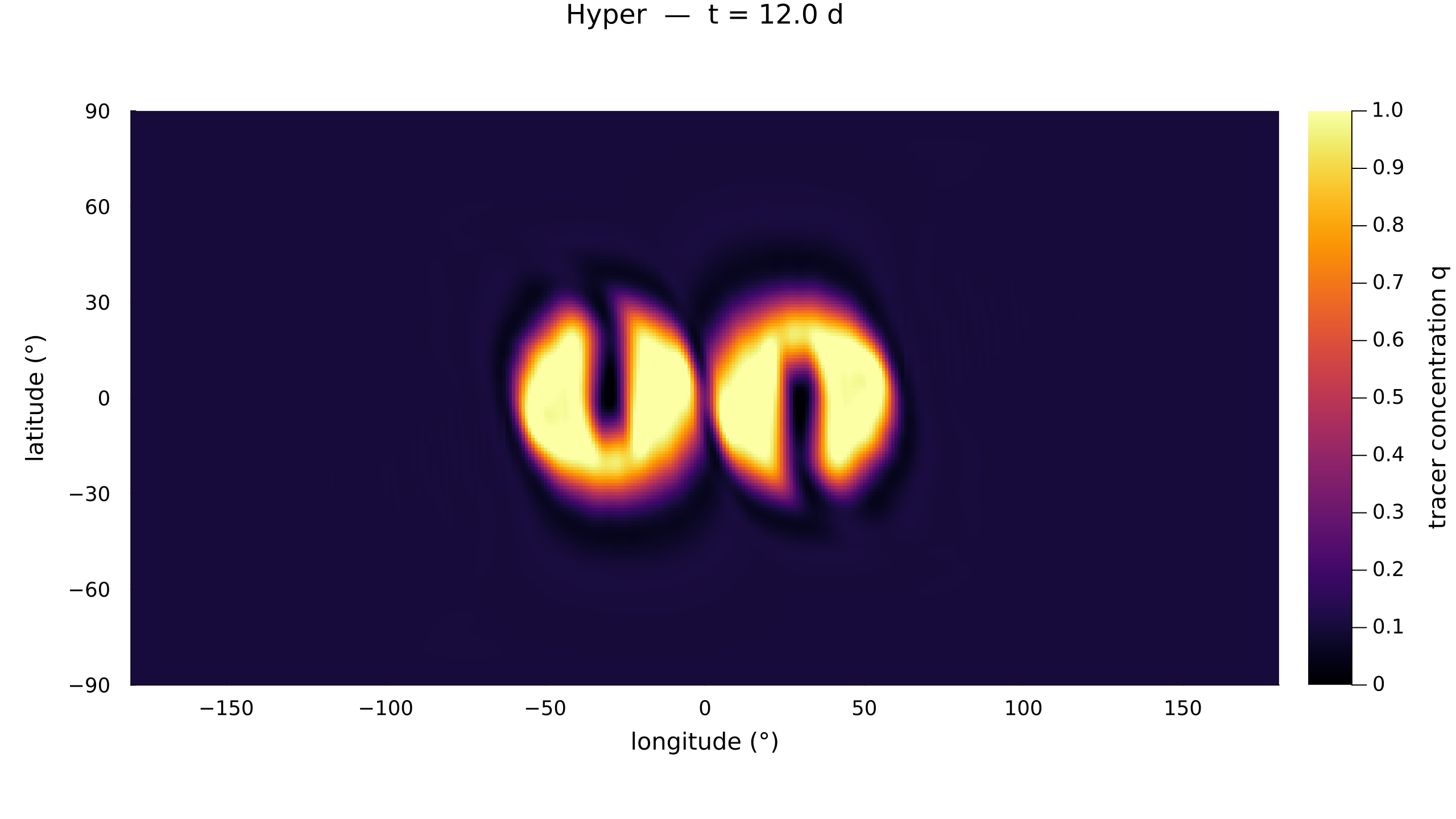}
        \caption{Hyperdiffusion.}
        \label{fig:hyperdiffusion-final}
    \end{subfigure}

    \caption{Tracer fields at final time point $t=T$ for four stabilization methods implemented in \texttt{ClimaCore.jl}.}
    \label{fig:stabilization-comparison}
\end{figure}

In Figure \ref{fig:stabilization-comparison} we show the tracer field at the final time point computed using \texttt{ClimaCore.jl} for four stabilization methods, including the case without stabilization\footnote{For completeness, the tracer fields  at both $t=T/2$ and $t=T$ for all ten ClimaCore.jl stabilization configurations are shown in Appendix \ref{app:all-stabilization-fields}.}. Observe that the solution without stabilization is able to retain the sharp features of the slotted cylinders quite well and the solution with SUPG is not visually different than that with no stabilization. As expected, we see more blurring/smearing around the sharp features of the cylinders in the case of the quasi-monotone limiter and even more in the hyperdiffusion case.

\noindent
Now turning our attention to Table \ref{tab:stabilization-comparison}, we can better assess the methods' performance in preventing unphysical extrema while retaining the overall tracer field. Observe that compared to the no stabilization case, the quasi-monotone limiter drastically reduces final-time overshoot and undershoot, with $q_{\mathrm{under}}$ and $q_{\mathrm{over}}$ reducing down to $10^{-7}–10^{-8}$. However, this is done at the expense of significantly larger global errors. Hyperdiffusion performs poorly across all of the computed metrics compared to the no stabilization case. Among the ClimaCore.jl stabilized solutions, the SUPG method performs best at preserving the tracer structure (as is evident by the smallest $\ell_1$, $\ell_2$, and $\ell_\infty$ errors), however it does not improve on the unstabilized baseline. Finally, in the cases where the limiter is included in a combined method, the extrema errors remain in the same $10^{-7}$--$10^{-8}$ range, showing that the behavior is analogous to that when the limiter method used alone. 

Overall, these results illustrate the tradeoff between preserving global structure and suppressing local overshoots and undershoots. The quasi-monotone limiter is most effective at preventing spurious oscillation, but does so at the expense of global errors. The SUPG method is better than the limiters at preserving the global structure, but has larger $q_{\text{under}}$ and $q_{\text{over}}$ errors. No single stabilization method of those tested is best across all metrics -- and most notably, the SUPG method does not improve on the no stabilization baseline for any of the metrics. 

It should be noted that the metrics chosen here for the definition of the errors, namely equations (\ref{eq:extrema_errors}) and (\ref{eq:global_errors}), only account for the difference between the final time solutions and the initial conditions. They do not track the temporal evolution over time. For a test case such as the one analyzed in this work, where the flow distorts the initial tracer configuration and then reverts it back to match its initial condition, tracking the history of the errors at every time step would give more information on the performance of the different stabilization methods in the transient stages. Finally, higher-order polynomial degree tests should be considered as they would highlight the benefits of the stabilization methods compared to the no stabilization results.

\section{Theoretical Analysis} \label{sec5}
\subsection{Conservation, boundedness, and variation control}

The exact transport system possesses two different kinds of structure. The
conservative variables $\rho$ and $Q=\rho q$ satisfy global conservation laws,
whereas the mixing ratio $q$ satisfies a pointwise transport equation. Indeed,
expanding the tracer-density equation gives
\begin{align}
0
&= \frac{\partial (\rho q)}{\partial t}
   + \nabla\cdot(\rho q\mathbf{u}) \\
&= \rho\left(\frac{\partial q}{\partial t}
   + \mathbf{u}\cdot\nabla q\right)
   + q\left(\frac{\partial\rho}{\partial t}
   + \nabla\cdot(\rho\mathbf{u})\right).
\end{align}
Consequently, wherever $\rho>0$, the density equation implies
\begin{equation}
    \frac{\partial q}{\partial t}+\mathbf{u}\cdot\nabla q=0.
    \label{eq:q_characteristic_transport}
\end{equation}
Thus $q$ is constant along characteristics of the exact velocity field. In
particular, the continuous solution satisfies
\begin{equation}
    \min_{\Omega} q(\cdot,0)
    \le q(\mathbf{x},t)
    \le \max_{\Omega} q(\cdot,0)
    \qquad \text{for all }t
    \label{eq:continuous_bounds}
\end{equation}
for which the characteristic flow remains well defined. For the present
slotted-cylinder problem, the physically relevant interval is therefore
$0.1\le q\le 1$.

\subsubsection*{Semidiscrete conservation}

The conservative SUPG formulation in \eqref{eq:supg_weak} preserves total
tracer mass independently of whether it preserves the pointwise bounds in
\eqref{eq:continuous_bounds}. To see this, choose the constant test function
$w_h=1$. This choice is admissible because the computational domain is the
closed sphere and the finite element space contains constants. Since
$\nabla 1=0$, both the advective weak term and the SUPG correction vanish.
There is also no boundary-flux contribution on a closed surface. Hence,
\begin{equation}
    \left(\frac{\partial Q_h}{\partial t},1\right)=0,
\end{equation}
which gives
\begin{equation}
    \frac{\mathrm{d}}{\mathrm{d}t}
    \int_{\Omega}Q_h\,\mathrm{d}\Omega=0.
    \label{eq:tracer_mass_conservation}
\end{equation}
The same argument applied to the density equation yields
\begin{equation}
    \frac{\mathrm{d}}{\mathrm{d}t}
    \int_{\Omega}\rho_h\,\mathrm{d}\Omega=0.
    \label{eq:density_mass_conservation}
\end{equation}
These identities hold at the semidiscrete level, up to quadrature and linear
solver tolerances. A Runge--Kutta method also preserves these linear
invariants provided that the residual at every stage has zero integral.

The factor $\rho_h$ in the tracer SUPG correction is nevertheless important.
It makes the stabilization act on the advective residual of the mixing ratio
while the prognostic variable $Q_h=\rho_h q_h$ remains conservative. This
mass-weighted construction supports consistency between the density and
tracer equations, including preservation of a spatially constant mixing
ratio. It does \emph{not}, by itself, imply $q_h\ge 0$ or prevent the creation
of new extrema.

\subsubsection*{Why standard SUPG is not bound preserving or TVD}

The SUPG term adds residual-based dissipation primarily in the streamline
direction \cite{brooks1982streamline}. This improves stability for
advection-dominated problems, but the resulting high-order finite element
operator is not generally monotone and does not satisfy a discrete maximum
principle. In particular, the consistent mass matrix and the off-diagonal
entries of the transport operator need not have the sign structure required
for a positivity-preserving update \cite{guermond2017mass}. This distinction
is visible in Table \ref{tab:stabilization-comparison}: the SUPG solution
conserves the transported quantity but still produces nonzero overshoot and
undershoot metrics, whereas the quasi-monotone limiter reduces both metrics
to nearly machine precision.

This limitation is consistent with the classical order barrier identified by
Godunov: within the usual linear one-step setting, a method that is monotone
for linear advection cannot be more than first-order accurate
\cite{godunov1959difference}. The theorem does not apply verbatim to every
multidimensional finite element formulation, but it identifies the basic
reason that a consistent residual-based stabilization such as SUPG (which is linear when applied to a linear PDE, and nonlinear otherwise) cannot
simultaneously provide sharp resolution and unconditional monotonicity. A
nonlinear limiting or nonlinear diffusion mechanism is required.

For a one-dimensional grid, the discrete total variation is commonly defined
by
\begin{equation}
    \mathrm{TV}(q^n)=\sum_j |q_{j+1}^n-q_j^n|,
\end{equation}
and a method is TVD if
$\mathrm{TV}(q^{n+1})\le \mathrm{TV}(q^n)$ \cite{harten1983high}. Strict TVD
should be interpreted cautiously for the present spherical deformational-flow
benchmark. Even the exact multidimensional transport map can stretch
interfaces and increase $\int_{\Omega}|\nabla q|\,\mathrm{d}\Omega$ during the
cycle. The more appropriate requirements here are preservation of the
physical interval $[0.1,1]$, absence of numerically generated extrema, and
control of excess variation relative to the exact transported field. Since
the flow returns the tracer to its initial configuration at $t=T$, the ratio
$\mathrm{TV}(q_h(T))/\mathrm{TV}(q_h(0))$ can still be used as an end-of-cycle
diagnostic, but a per-step TVD inequality is not expected from the exact
multidimensional dynamics.

\subsubsection*{Modifications needed for sharper and bound-preserving transport}

A first extension is a residual-dependent discontinuity-capturing or
crosswind-diffusion term. One representative form is
\begin{equation}
\mathcal{S}_{\mathrm{DC}}(q_h,w_h)
=
\sum_{K}\int_K
\nu_{\mathrm{DC},K}
\left(\mathbf{P}_{\perp}\nabla q_h\right)\cdot
\left(\mathbf{P}_{\perp}\nabla w_h\right)
\,\mathrm{d}\Omega,
\label{eq:dc_term}
\end{equation}
where
\begin{equation}
    \mathbf{P}_{\perp}
    =\mathbf{I}-\widehat{\mathbf{u}}\otimes\widehat{\mathbf{u}},
    \qquad
    \widehat{\mathbf{u}}=\frac{\mathbf{u}}{\|\mathbf{u}\|},
\end{equation}
and $\nu_{\mathrm{DC},K}\ge0$ is activated by a local residual or gradient
sensor. At points where $\|\mathbf{u}\|$ is zero or sufficiently small, the
projector must be regularized, for example by replacing
$\|\mathbf{u}\|$ with $\max(\|\mathbf{u}\|,\varepsilon_u)$ for a small
positive tolerance $\varepsilon_u$. The original ``beyond SUPG'' construction adds precisely this type of discontinuity-capturing mechanism to control strong internal and boundary
layers \cite{hughes1986new}. Such a term damps crosswind oscillations that
streamline diffusion alone cannot see and can substantially reduce ringing
near the slots. However, a generic shock-capturing term is not automatically
TVD or maximum-principle preserving; its coefficient and nonlinear sensor
must be designed with those properties in mind.

A stronger route is conservative algebraic flux correction or convex
limiting. In algebraic form, write the high-order semidiscrete method as
\begin{equation}
    \mathbf{M}_{C}\dot{\mathbf{q}}
    +\mathbf{K}_{H}\mathbf{q}=0,
\end{equation}
where $\mathbf{M}_{C}$ is the consistent mass matrix and
$\mathbf{K}_{H}$ contains the Galerkin and SUPG contributions. A
bound-preserving low-order method is first constructed using a lumped mass
matrix $\mathbf{M}_{L}$ and sufficient graph viscosity:
\begin{equation}
    \mathbf{M}_{L}\dot{\mathbf{q}}
    +\mathbf{K}_{L}\mathbf{q}=0.
\end{equation}
The low-order update is chosen to satisfy a local maximum principle under an
appropriate CFL restriction. The difference between the high- and low-order
operators is then decomposed into pairwise antidiffusive fluxes
$f_{ij}=-f_{ji}$. Replacing each flux by
$\alpha_{ij}f_{ij}$, with $0\le\alpha_{ij}\le1$, restores as much high-order
accuracy as the local bounds permit while retaining conservation. This is
the finite element flux-correction framework developed in
\cite{kuzmin2002flux}; related invariant-domain and maximum-principle
preserving continuous finite element constructions are given in
\cite{guermond2017invariant}.

For the coupled density--tracer system, the conservative variable $Q_h$ should
continue to be updated in mass form, but the admissible bounds should be
imposed on $q_h=Q_h/\rho_h$. The density and tracer fluxes must therefore be
limited consistently. Pointwise clipping of $q_h$ after a time step would
restore nonnegativity but generally destroy tracer-mass conservation, whereas
pairwise conservative flux limiting can enforce the bounds and preserve
\eqref{eq:tracer_mass_conservation}. When an SSP Runge--Kutta method is used,
the limiter must be applied at every forward-Euler stage so that the fully
discrete update remains a convex combination of bound-preserving stages
\cite{shu1988efficient}.

These additions have clear benefits: they prevent physically impossible
negative tracer concentrations, suppress spurious ringing, and make the
extrema metrics in \eqref{eq:extrema_errors} controlled properties of the
algorithm rather than empirical outcomes. The tradeoffs are additional
nonlinearity, a CFL restriction for explicit bound preservation, and some
extra diffusion near discontinuities. A useful practical hierarchy is
therefore SUPG for streamline stability, residual-based discontinuity
capturing for unresolved crosswind gradients, and conservative flux limiting
only where the physical bounds would otherwise be violated.

\subsection{Asymptotic Approach to Time-Dependent SUPG}
In Section \ref{sec3}, we explored the efficacy of the time-independent SUPG implementation using the stabilizing term
\[
    \tau^e\int_\Omega (u\cdot\nabla v)\left(\nabla\cdot (Qu)+\frac{\partial Q}{\partial t}\right)
\]
Where the term $\frac{\partial Q}{\partial t}$ is computed from the \textit{previous} time step as a proxy for the $\frac{\partial Q}{\partial t}$ field at the current time step. While this allows us to set up an explicit method rather than an implicit method, it suffers from inaccuracy for large time steps. Rather than pursuing a computationally costly implicit SUPG method, we take an asymptotic expansion based approach to improve our proxy for $\frac{\partial Q}{\partial t}$ at the current time step.

Consider the SUPG stabilizing term (assuming no source, i.e. $s=0$)
\[
\tau^e\int_\Omega(u\cdot\nabla v)\left(\nabla\cdot (Qu)+\frac{\partial Q}{\partial t}\right)
\]
\[
=\tau^e\int_\Omega\nabla v\cdot \left[u \left(\nabla\cdot (Qu)+\frac{\partial Q}{\partial t}\right)\right]
\]
And since $\nabla\cdot(AB)=(\nabla A)\cdot B+A(\nabla\cdot B)$, this can be written as
\[
=\tau^e\int_\Omega\nabla\cdot \left\{v \left[u\left(\nabla\cdot (Qu)+\frac{\partial Q}{\partial t}\right)\right]\right\}-\tau^e\int_\Omega v\left\{\nabla\cdot \left[u\left(\nabla\cdot (Qu)+\frac{\partial Q}{\partial t}\right)\right]\right\}
\]
By the divergence theorem,
\[
\tau^e\int_\Omega\nabla\cdot \left\{v\left[u\left(\nabla\cdot (Qu)+\frac{\partial Q}{\partial t}\right)\right]\right\}=\tau^e\oint_{\partial\Omega}
v\left[u\left(\nabla\cdot (Qu)+\frac{\partial Q}{\partial t}\right)\right]\cdot\hat{\mathbf{n}}=0,
\]
since $v$ vanishes on $\partial\Omega$. Thus, the SUPG stabilizing term is
\[
-\tau^e\int_\Omega v\left\{\nabla\cdot \left[u\left(\nabla\cdot (Qu)+\frac{\partial Q}{\partial t}\right)\right]\right\}
\]
Then the tracer dynamics are described by
\[
\int_{\Omega}v\left(\frac{\partial Q}{\partial t}\right)=-\int_{\Omega}v(\nabla\cdot(Qu))-\tau^e\int_\Omega v\left\{\nabla\cdot \left[u\left(\nabla\cdot (Qu)+\frac{\partial Q}{\partial t}\right)\right]\right\}.
\]
We now perform a nondimensionalization of this problem. Take
\[
Q=c\bar{Q},\quad u=U\bar{u},\quad t=T\bar{t},\quad \nabla\cdot=\frac{1}{L}\bar{\nabla}\cdot
\]
so we have
\begin{align*}
   \frac{c}{T}\int_{\Omega}v\left(\frac{\partial \bar{Q}}{\partial \bar{t}}\right)
   =&-\frac{cU}{L}\int_{\Omega}v(\bar{\nabla}\cdot(\bar{Q}\bar{u}))-\tau^e\frac{cU^2}{L^2}\int_\Omega v\{\bar{\nabla}\cdot [\bar{u}(\bar{\nabla}\cdot(\bar{Q}\bar{u})]\}\\
    &-\tau^e\frac{cU}{LT}\int_{\Omega}v\left\{\bar{\nabla}\cdot\left[\bar{u}\left(\frac{\partial \bar{Q}}{\partial \bar{t}}\right)\right]\right\}. 
\end{align*}

By the substitution $T=L/U$, we get
\begin{align*}
    \int_{\Omega}v\left(\frac{\partial \bar{Q}}{\partial \bar{t}}\right)
    = &-\int_{\Omega}v(\bar{\nabla}\cdot(\bar{Q}\bar{u}))-\tau^e\frac{U}{L}\int_\Omega v\{\bar{\nabla}\cdot [\bar{u}(\bar{\nabla}\cdot(\bar{Q}\bar{u})]\}\\
    &-\tau^e\frac{U}{L}\int_{\Omega}v\left\{\bar{\nabla}\cdot\left[\bar{u}\left(\frac{\partial \bar{Q}}{\partial \bar{t}}\right)\right]\right\}
\end{align*}

\begin{figure}[H]
\begin{center}
\begin{tikzpicture}[
  node distance=3cm,
  every node/.style={font=\large},
  arr/.style={-{Stealth[length=8pt]}, thick}
]

\node (TL) {$Q_0^{n}$};
\node (TR) [right=3cm of TL] {$Q_1^{n}$};
\node (BL) [below=of TL] {$Q_0^{n+1}$};
\node (BM) [right=3cm of BL] {$\dfrac{\partial Q_i^{n+1}}{\partial t}$};
\node (BR) [right=3cm of BM] {$Q_1^{n+1}$};

\draw[arr] (TL) -- (BL) node[midway, left] {$$\boxed{1}$$};
\draw[arr] (TL) -- (BM) node[midway, midway, xshift=20pt] {$$\boxed{2}$$};
\draw[arr] (BL) -- (BM) node[above, midway] {$$\boxed{2}$$};
\draw[arr] (BM) -- (BR) node[above, midway] {$$\boxed{3}$$};
\draw[arr] (TR) -- (BR) node[midway, midway, xshift=20pt] {$$\boxed{3}$$};

\draw[arr] (BL) to[bend right=30] node[above, pos=0.5] {$\boxed{3}$} (BR);
\end{tikzpicture}
\end{center}
\caption{Diagrammatic representation of the solution algorithm for the asymptotic approach of the pure advection problem with SU stabilization.}
\label{fig:asymptotic_fig}
\end{figure}
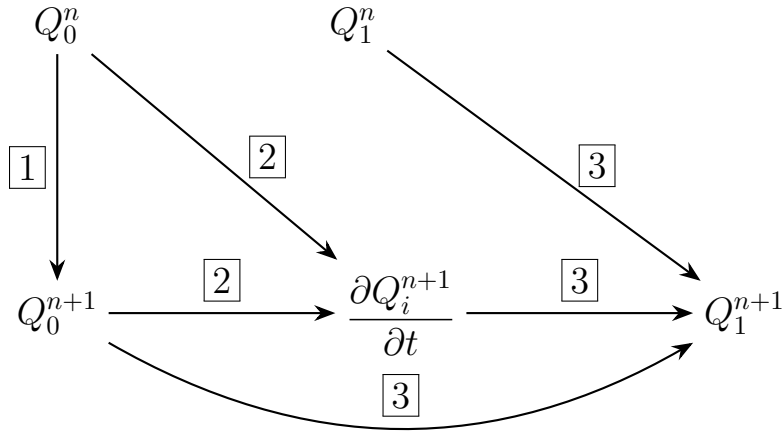

In our atmospheric simulation of the Earth, $U\approx10^2[km/hr]$, $L\approx6\cdot10^3[km]$. We have $\tau^e\approx2.5\cdot10^{-1}[hr]$. Thus,
\[
\tau^e\frac{U}{L}\approx4\cdot10^{-3}\ll1.
\]
Letting $\varepsilon:=\tau^e\frac{U}{L}$, we have
\begin{equation}
    \begin{aligned}
        \int_{\Omega}v\left(\frac{\partial \bar{Q}}{\partial \bar{t}}\right) &= -\int_{\Omega}v(\bar{\nabla}\cdot(\bar{Q}\bar{u})) \\
        &\phantom{=\,} -\varepsilon\int_\Omega v(\bar{\nabla}\cdot \{\bar{u}[\bar{\nabla}\cdot(\bar{Q}\bar{u})]\})-\varepsilon\int_{\Omega}v\left\{\bar{\nabla}\cdot\left[\bar{u}\left(\frac{\partial \bar{Q}}{\partial \bar{t}}\right)\right]\right\}
    \end{aligned}
\end{equation}
From this point on, we will neglect the overbar notation (e.g. $\bar{Q}$) in the non-dimensionalization for simplicity. Consider the asymptotic expansion
\[
Q=Q_0+\varepsilon Q_1+...
\]
Substituting into ($19$) and expanding up to $\mathcal{O}(1)$, we get the leading term relation
\begin{align} \label{standardG}
    \int_{\Omega}v\left(\frac{\partial Q_0}{\partial t}\right)=-\int_{\Omega}v(\nabla\cdot(Q_0u))
\end{align}
Expanding up to $\mathcal{O}(\varepsilon)$, we have
\begin{equation} \label{eq:leading_term}
\begin{aligned}
    \int_{\Omega}v\left(\frac{\partial Q_1}{\partial t}\right)&=-\int_{\Omega}v(\nabla\cdot(Q_1u))-\int_\Omega v\{\nabla\cdot [u(\nabla\cdot(Q_0u))]\} \\
    &\phantom{=\,}-\int_{\Omega}v\left\{\nabla\cdot \left[u\left(\frac{\partial Q_0}{\partial t}\right)\right]\right\}.
\end{aligned}
\end{equation}
If we consider the time discretization
\begin{align} \label{eq:t_discretization}
    \frac{\partial Q_i^{n}}{\partial t}\approx\frac{Q_i^{n}-Q_i^{n-1}}{\Delta t},
\end{align}
we can construct an algorithm based on solving consecutive pairs $(Q_0^{n+1},Q_1^{n+1})$ from $(Q_0^n,Q_1^n)$. First, we initialize the start system based on the initial concentration $Q$. We can set 
\[
(Q_0^0,Q_1^0)=(Q,0)
\]
Then, given some pair $(Q_0^n,Q_1^n)$ at the $n^{th}$ time step, we follow the diagram in Figure \ref{fig:asymptotic_fig} and the algorithm outlined below.

\noindent
\begin{enumerate}
    \item Use the RK4 algorithm to compute $Q_0^{n+1}$ from $Q_0^n$ within the standard Galerkin method in (\ref{standardG})
    \item Numerically compute $\frac{\partial Q_0^{n+1}}{\partial t}$ from $Q_0^n$ and $Q_0^{n+1}$ in (\ref{eq:t_discretization}). 
    \item Use the SU method to compute $Q_1^{n+1}$ from $Q_1^n$, $Q_0^{n+1}$, and $\frac{\partial Q_0^{n+1}}{\partial t}$ in (\ref{eq:leading_term}). 
    \item Set $n=n+1$, compute the tracer concentration $Q^n=Q_0^n+\varepsilon Q_1^n$, and repeat from step 1.
\end{enumerate}
\noindent
While this algorithm involves solving two discrete Galerkin-type problems at each time step (standard Galerkin for $Q_0^{n+1}$, SU for $Q_1^{n+1}$), it allows us to include a non-delayed time-dependent term in the strong-form residual in the stabilizing term.

It should be noted that our asymptotic expansion $Q=Q_0+\varepsilon Q_1$ is arbitrary, and could contain more terms. The algorithm detailed in this section is but a member of a class of algorithms; the algorithm where the number of expansion terms in $n=2$. For example, with the addition of a third term $Q_2$ scaled by $\varepsilon^2$, we could detail an algorithm which requires solving three discrete Galerkin-type problems at each time step. Computation scales linearly with the number of expansion terms, as $n$ terms requires $n$ Galerkin-type problems to be solved at each time step. The choice for the number of expansion terms used should be based on comparisons with implicit SUPG methods, as well as computational constraints.

\section{Conclusions and Future Work}

In this work, we have implemented the Streamline Upwind (SU) and Streamline Upwind Petrov-Galerkin (SUPG) stabilization methods for finite element methods (FEM) using the \texttt{ClimaCore.jl} library \cite{ClimaCore_github}. An implementation for SU is also developed in Firedrake \cite{FiredrakeUserManual} to compare and contrast it with the implementation in ClimaCore.jl. Other stabilization methods, such as hyperdiffusion and flux-limiters, are also implemented in ClimaCore.jl and we conduct a cross-comparison of these four stabilization methods. 

The comparison between the stabilization methods was performed using the slotted-cylinders deformation flow benchmark problem. For this problem, we found that the solution obtained without numerical stabilization is able to resolve the sharp features of the cylinders quite well and gave the smallest global errors. Stabilization using quasi-monotone limiters greatly improved at reducing spurious extrema at the expense of global accuracy, leading to smearing over the course of the simulation. The hyperdiffusion method performed relatively poorly across all metrics. When methods were combined, the flux-limiter method was shown to largely determine the extrema behavior, with hyperdiffusion, SU, and SUPG having minimal additive impact. In a comparison of implementation, both Python's Firedrake and Julia's ClimaCore.jl produced errors at the same order of magnitude, with the notable exception of the $\ell_\infty$-error, where they differ by a single order of magnitude. The SUPG method was not found to improve upon the no stabilization baseline, for the metrics tested in this work. However, among the stabilized runs, the SUPG method resulted in the lowest global errors. 

Following these results, multiple directions for this research may be pursued. For instance, additional test cases with a variety of tracer configurations and fluid regimes would provide a more comprehensive comparison of the stabilization methods. Moreover, additional error tracking measuring the error over time (rather than only at the final time) would give a more complete view. Finally, our implementations of the the SU and SUPG methods would benefit from different sets of parameter studies. 


The theoretical analysis also clarifies that conservation and bound preservation are distinct properties. The conservative semidiscrete SUPG formulation preserves total density and tracer mass on the closed spherical domain, but standard linear SUPG is not generally maximum-principle preserving or total variation diminishing. The overshoots and undershoots observed numerically are therefore consistent with the structure of the method. A future bound-preserving implementation could combine SUPG with residual-based crosswind discontinuity capturing and a conservative flux or convex limiter applied consistently to the density and tracer fluxes.

An asymptotics based SUPG algorithm based on the SU and delayed SUPG methods was developed, but remains to be implemented. For potential future research, this algorithm could be implemented and compared with implicit SUPG methods, and the impact of the number of asymptotic expansion terms on accuracy could be studied.

\section{Acknowledgments}

We would like to gratefully acknowledge the support of SIAM and the University of Delaware for providing the resources to conduct this work. Additionally, a special thank you to Dr. Valeria Barra for her excellent mentorship and support. 

\section{AI Use Disclosure}
The code development for the Firedrake implementation was aided by Claude Opus 4.8.

\printbibliography

@article{nair2010class,
  title={A class of deformational flow test cases for linear transport problems on the sphere},
  author={Ramachandran D. Nair and Peter H. Lauritzen},
  journal={Journal of Computational Physics},
  year={2010},
  publisher={Elsevier},
  doi={https://doi.org/10.1016/j.jcp.2010.08.014}
}

@article{brooks1982streamline,
  title={Streamline upwind/Petrov-Galerkin formulations for convection dominated flows with particular emphasis on the incompressible Navier-Stokes equations},
  author={Alexander N. Brooks and Thomas J.R. Hughes},
  journal={Computer Methods in Applied Mechanics and Engineering},
  year={1982},
  publisher={North-Holland},
  doi={https://doi.org/10.1016/0045-7825(82)90071-8}
  }

@article{guba2014optimization,
  title={Optimization-based limiters for the spectral element method},
  author={Oksana Guba and Mark Taylor and Amik St-Cyr},
  journal={Journal of Computational Physics},
  year={2010},
  publisher={Elsevier},
  doi={https://doi.org/10.1016/j.jcp.2014.02.029}
  }

@article{hughes1986new,
  title={A new finite element formulation for computational fluid dynamics: II. Beyond SUPG},
  author={Thomas J.R. Hughes, Michel Mallet and Akira Mizukami},
  journal={Computer Methods in Applied Mechanics and Engineering},
  year={1986},
  publisher={North-Holland},
  doi={https://doi.org/10.1016/0045-7825(86)90110-6}
  }

@article{tezduyar2003stabilization,
  title={Stabilization Parameters in SUPG and PSPG Formulations},
  author={Tayfun Tezduyar and Sunil Sathe},
  journal={Journal of Computational and Applied Mathematics},
  year={2003},
  publisher={Elsevier}
  }

@article{shu1988efficient,
  title={Efficient implementation of essentially non-oscillatory shock-capturing schemes},
  author={Shu, Chi-Wang and Osher, Stanley},
  journal={Journal of computational physics},
  %volume={77},
  %number={2},
  %pages={439--471},
  year={1988},
  publisher={Elsevier},
  doi={https://doi.org/10.1016/0021-9991(88)90177-5}
}

@software{ClimaCore_github,
  author={CliMA},
  title={ClimaCore.jl},
  month={04},
  year={2025},
  publisher={GitHub},
  version={v0.14.51},
  url={https://github.com/CliMA/ClimaCore.jl.git}
  }

@article{ClimaCore_dycore_paper,
    author = {Yatunin, Dennis and Byrne, Simon and Kawczynski, Charles and Kandala, Sriharsha and Bozzola, Gabriele and Sridhar, Akshay and Shen, Zhaoyi and Jaruga, Anna and Sloan, Julia and He, Jia and Huang, Daniel Zhengyu and Barra, Valeria and Chew, Ray and Boral, Anudhyan and Chen, Yi-fan and Knoth, Oswald and Ullrich, Paul and Mbengue, Cheikh and Schneider, Tapio},
    title = {The Climate Modeling Alliance Atmosphere Dynamical Core: Concepts, Numerics, and Scaling},
    journal = {Journal of Advances in Modeling Earth Systems},
    volume = {18},
    number = {3},
    pages = {e2025MS005014},
    doi = {https://doi.org/10.1029/2025MS005014},
    url = {https://agupubs.onlinelibrary.wiley.com/doi/abs/10.1029/2025MS005014},
    year = {2026}
}

@article{godunov1959difference,
  author  = {Godunov, Sergei K.},
  title   = {A Difference Method for Numerical Calculation of Discontinuous Solutions of the Equations of Hydrodynamics},
  journal = {Matematicheskii Sbornik},
  volume  = {47},
  number  = {3},
  pages   = {271--306},
  year    = {1959},
  note    = {In Russian; English translation published as U.S. Joint Publications Research Service report JPRS 7226 (1969)}
}

@article{harten1983high,
  author  = {Harten, Ami},
  title   = {High Resolution Schemes for Hyperbolic Conservation Laws},
  journal = {Journal of Computational Physics},
  volume  = {49},
  number  = {3},
  pages   = {357--393},
  year    = {1983},
  doi     = {10.1016/0021-9991(83)90136-5}
}

@article{kuzmin2002flux,
  author  = {Kuzmin, Dmitri and Turek, Stefan},
  title   = {Flux Correction Tools for Finite Elements},
  journal = {Journal of Computational Physics},
  volume  = {175},
  number  = {2},
  pages   = {525--558},
  year    = {2002},
  doi     = {10.1006/jcph.2001.6955}
}

@article{guermond2017invariant,
  author  = {Guermond, Jean-Luc and Popov, Bojan},
  title   = {Invariant Domains and Second-Order Continuous Finite Element Approximation for Scalar Conservation Equations},
  journal = {SIAM Journal on Numerical Analysis},
  volume  = {55},
  number  = {6},
  pages   = {3120--3146},
  year    = {2017},
  doi     = {10.1137/16M1106560}
}

@article{guermond2017mass,
  author  = {Guermond, Jean-Luc and Popov, Bojan and Yang, Yong},
  title   = {The Effect of the Consistent Mass Matrix on the Maximum-Principle for Scalar Conservation Equations},
  journal = {Journal of Scientific Computing},
  volume  = {70},
  number  = {3},
  pages   = {1358--1366},
  year    = {2017},
  doi     = {10.1007/s10915-016-0285-7}
}

@manual{FiredrakeUserManual,
  title        = {Firedrake User Manual},
  author       = {David A. Ham and Paul H. J. Kelly and Lawrence Mitchell and Colin J. Cotter and Robert C. Kirby and Koki Sagiyama and Nacime Bouziani and Sophia Vorderwuelbecke and Thomas J. Gregory and Jack Betteridge and Daniel R. Shapero and Reuben W. Nixon-Hill and Connor J. Ward and Patrick E. Farrell and Pablo D. Brubeck and India Marsden and Thomas H. Gibson and Miklós Homolya and Tianjiao Sun and Andrew T. T. McRae and Fabio Luporini and Alastair Gregory and Michael Lange and Simon W. Funke and Florian Rathgeber and Gheorghe-Teodor Bercea and Graham R. Markall},
  organization = {Imperial College London and University of Oxford and Baylor University and University of Washington},
  edition      = {First edition},
  year         = {2023},
  month        = {5},
  doi          = {10.25561/104839},
}

@misc{gusto_software,
  author       = {{The Gusto Development Team}},
  title        = {Gusto: A library for dynamical cores using compatible finite element discretisations},
  year         = {2026},
  publisher    = {GitHub},
  journal      = {GitHub repository},
  howpublished = {\url{https://github.com/firedrakeproject/gusto}}
}

@article{Ullrich2018,
   author = {Paul A. Ullrich and Daniel R. Reynolds and Jorge E. Guerra and Mark A. Taylor},
   doi = {10.1016/j.jcp.2018.06.035},
   issn = {10902716},
   journal = {Journal of Computational Physics},
   month = {12},
   pages = {427-446},
   title = {Impact and importance of hyperdiffusion on the spectral element method: A linear dispersion analysis},
   volume = {375},
   year = {2018}
}

@article{Zalesak1979,
    title = {Fully multidimensional flux-corrected transport algorithms for fluids},
    journal = {Journal of Computational Physics},
    volume = {31},
    number = {3},
    pages = {335-362},
    year = {1979},
    issn = {0021-9991},
    doi = {https://doi.org/10.1016/0021-9991(79)90051-2},
    url = {https://www.sciencedirect.com/science/article/pii/0021999179900512},
    author = {Steven T Zalesak}
}

\newpage
\appendix

\section{Tracer Fields for All Stabilization Configurations}
\label{app:all-stabilization-fields}

\newlength{\appendixpanelheight}
\setlength{\appendixpanelheight}{0.135\textheight}

\newlength{\appendixpanelsidetrim}
\setlength{\appendixpanelsidetrim}{0cm}

\newlength{\appendixpanelbottomtrim}
\setlength{\appendixpanelbottomtrim}{6cm}

\newlength{\appendixpaneltoptrim}
\setlength{\appendixpaneltoptrim}{4cm}

\newcommand{\methodpanel}[3]{%
    \begin{subfigure}[b]{0.48\textwidth}
        \centering
        \begingroup
        \edef\appendixincludegraphics{%
            \endgroup
            \noexpand\includegraphics[
                height=\the\appendixpanelheight,
                keepaspectratio,
                trim=\the\appendixpanelsidetrim\space
                     \the\appendixpanelbottomtrim\space
                     \the\appendixpanelsidetrim\space
                     \the\appendixpaneltoptrim,
                clip
            ]{animations/#1/frames/#2}%
        }%
        \appendixincludegraphics
        \caption{#3}
    \end{subfigure}%
}

\begin{figure}[htbp!]
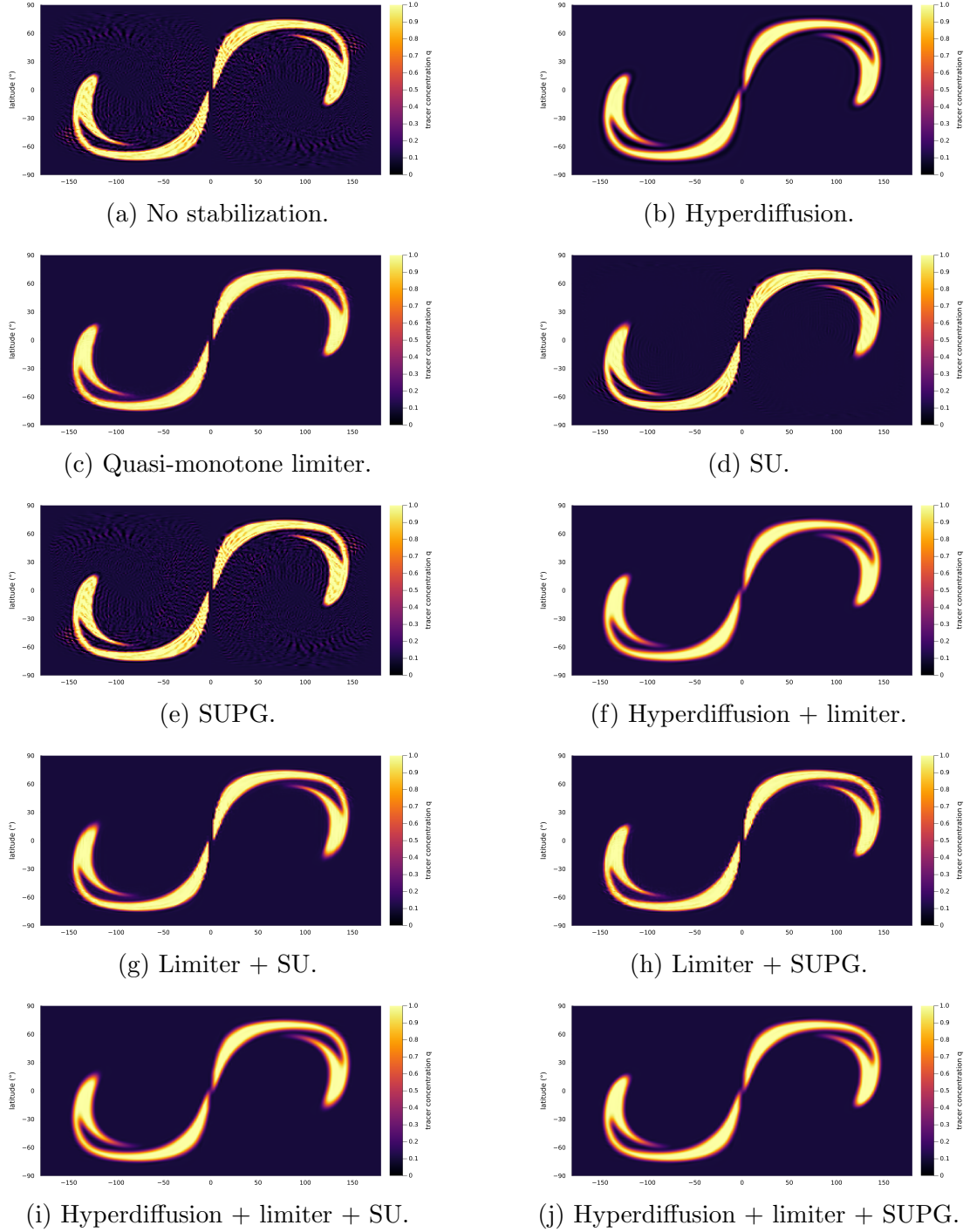

    \centering

    \methodpanel{None}{frame_060.png}{No stabilization.}
    \hfill
    \methodpanel{Hyper}{frame_060.png}{Hyperdiffusion.}
    \par\vspace{0.3em}

    \methodpanel{Limiter}{frame_060.png}{Quasi-monotone limiter.}
    \hfill
    \methodpanel{SU}{frame_060.png}{SU.}
    \par\vspace{0.3em}

    \methodpanel{SUPG}{frame_060.png}{SUPG.}
    \hfill
    \methodpanel{Hyper_Limiter}{frame_060.png}{Hyperdiffusion + limiter.}
    \par\vspace{0.3em}

    \methodpanel{Limiter_SU}{frame_060.png}{Limiter + SU.}
    \hfill
    \methodpanel{Limiter_SUPG}{frame_060.png}{Limiter + SUPG.}
    \par\vspace{0.3em}

    \methodpanel{Hyper_Limiter_SU}{frame_060.png}
        {Hyperdiffusion + limiter + SU.}
    \hfill
    \methodpanel{Hyper_Limiter_SUPG}{frame_060.png}
        {Hyperdiffusion + limiter + SUPG.}

    \caption{Tracer fields at the intermediate time $t=T/2$ for all
    stabilization choices implemented in ClimaCore.jl.}
    \label{fig:all-methods-half}
\end{figure}

\begin{figure}[htbp!]
    \centering

    \methodpanel{None}{frame_119.png}{No stabilization.}
    \hfill
    \methodpanel{Hyper}{frame_119.png}{Hyperdiffusion.}
    \par\vspace{0.3em}

    \methodpanel{Limiter}{frame_119.png}{Quasi-monotone limiter.}
    \hfill
    \methodpanel{SU}{frame_119.png}{SU.}
    \par\vspace{0.3em}

    \methodpanel{SUPG}{frame_119.png}{SUPG.}
    \hfill
    \methodpanel{Hyper_Limiter}{frame_119.png}{Hyperdiffusion + limiter.}
    \par\vspace{0.3em}

    \methodpanel{Limiter_SU}{frame_119.png}{Limiter + SU.}
    \hfill
    \methodpanel{Limiter_SUPG}{frame_119.png}{Limiter + SUPG.}
    \par\vspace{0.3em}

    \methodpanel{Hyper_Limiter_SU}{frame_119.png}
        {Hyperdiffusion + limiter + SU.}
    \hfill
    \methodpanel{Hyper_Limiter_SUPG}{frame_119.png}
        {Hyperdiffusion + limiter + SUPG.}

    \caption{Tracer fields at the final time $t=T$ for all stabilization
    choices implemented in ClimaCore.jl.}
    \label{fig:all-methods-final}
\end{figure}
\end{document}